\documentclass[11pt,english,onecolumn]{IEEEtran}
\usepackage[T1]{fontenc}
\usepackage[latin9]{inputenc}
\usepackage{geometry}
\usepackage{amsmath}
\usepackage{amsthm}
\usepackage{amssymb}
\usepackage{graphicx}
\usepackage{setspace}
\makeatletter
\theoremstyle{plain}
\newtheorem{thm}{\protect\theoremname}
\theoremstyle{plain}
\newtheorem{lem}[thm]{\protect\lemmaname}
\theoremstyle{definition}
\newtheorem{example}[thm]{\protect\examplename}
\theoremstyle{plain}
\newtheorem{prop}[thm]{\protect\propositionname}
\theoremstyle{plain}
\newtheorem{fact}[thm]{\protect\factname}

\usepackage{hyperref}
\usepackage{enumitem}
\usepackage{breqn}
\usepackage{bbm} 
\usepackage{cite}
\allowdisplaybreaks

\renewcommand\[{\begin{equation}}
\renewcommand\]{\end{equation}}

\DeclareMathOperator*{\argmax}{arg\,max}

\global\long\def\P{\mathbb{P}}

\global\long\def\E{\mathbb{E}}
\global\long\def\I{\mathbbm{1}}
\global\long\def\d{\mathrm{d}}

\global\long\def\trre[#1,#2]{\overset{{\scriptstyle (#2)}}{#1}} % transition explained with reason

\author{Nir Weinberger$^{1}$ and Ilan Shomorony$^{2}$}

\makeatother

\usepackage{babel}
\providecommand{\examplename}{Example}
\providecommand{\factname}{Fact}
\providecommand{\lemmaname}{Lemma}
\providecommand{\propositionname}{Proposition}
\providecommand{\theoremname}{Theorem}

\begin{document}
\title{Fundamental Limits of Reference-Based\\
Sequence Reordering\thanks{$^{1}$ Nir Weinberger is with the Viterbi Faculty of Electrical and
Computer Engineering, Technion-Israel Institute of Technology, Haifa
3200004, Israel (Email: nirwein@technion.ac.il.). $^{2}$ Ilan Shomorony
is with the Department of Electrical and Computer Engineering, University
of Illinois at Urbana-Champaign, Urbana, IL 61801 USA (Email: ilans@illinois.edu).
This paper was presented in part in the 2023 IEEE International Symposium
on Information Theory. The research of N. W. was supported by the
Israel Science Foundation (ISF), grant no. 1782/22. The work of I.
S. was supported in part by the National Science Foundation under
CCF grants 2007597 and 2046991.}}

\maketitle

\renewcommand\[{\begin{equation}}
\renewcommand\]{\end{equation}}
\renewenvironment{align*}{\align}{\endalign}
\thispagestyle{empty}

\begin{abstract}
The problem of reconstructing a sequence of independent and identically
distributed symbols from a set of equal size, consecutive, fragments,
as well as a dependent reference sequence, is considered. First, in
the regime in which the fragments are relatively long, and typically
no fragment appears more than once, the scaling of the failure probability
of maximum likelihood reconstruction algorithm is exactly determined
for perfect reconstruction and bounded for partial reconstruction.
Second, the regime in which the fragments are relatively short and
repeating fragments abound is characterized. A trade-off is stated
between the fraction of fragments that cannot be adequately reconstructed
vs. the distortion level allowed for the reconstruction of each fragment,
while still allowing vanishing failure probability.
\end{abstract}


\section{Introduction}

\begin{figure}
\begin{centering}
\includegraphics{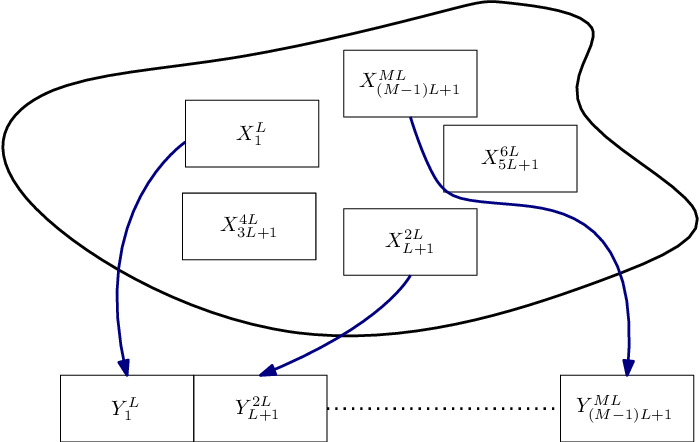}
\par\end{centering}
\caption{Illustration of the reference-based reordering problem \label{fig:Illustration of re-ordering problem}}

\end{figure}

In this paper, we consider the problem of reconstructing a sequence
$X^{N}\in{\cal X}^{N}$ from its non-overlapping consecutive fragments
and a reference sequence, as illustrated in Fig. \ref{fig:Illustration of re-ordering problem}.
A sequence of $N$ independent and identically distributed (IID) symbols
is drawn from a finite alphabet source, and is then partitioned into
non-overlapping, consecutive \emph{fragments} of length $L$ each.
The fragments are then permuted in an arbitrary manner, and a \emph{multiset}
of $M=N/L$ fragments is observed, without any specific order. In
order to facilitate the correct reordering of the fragments, the observer
of the fragments is supplied with a \emph{reference sequence} $Y^{N}\in{\cal Y}^{N}$
of length $N$. This reference sequence is similar, yet not identical,
to the sequence of interest; for example, it can be its noisy version,
or slightly different due to statistical variations in some population.
Concretely, it is assumed that each symbol in $Y^{N}$ is generated
by passing the corresponding symbol in a memoryless channel $P_{Y|X}$.
A reconstruction algorithm observes the $M$ fragments of the original
sequence as well as the reference sequence $Y^{N}$, and is required
to recover the original sequence $X^{N}$. 

This problem is motivated by settings in which data is observed out
of order, and ordering is made possible through side information.
It arises in various domains: First, DNA sequencing of a genomic sequence
typically produces short fragments, which should be assembled in order
to obtain the correct sequence.\footnote{More accurately, the fragments in DNA sequencing typically start at
random locations along the sequence, and might have overlaps. In this
sense, our model is a distilled version of this problem, and extending
our results to overlapping fragments is an interesting open problem. } Due to the high similarity between individual genomes, the reconstruction
algorithm may have access to a reference sequence, which can be used
to assemble the target sequence \cite{mohajer2013reference}. Second,
as described in \cite{kovavcevic2018codes}, such problem arises in
transmission of information over (noiseless) permutation channels,
such as packet networks employing multipath routing as a means for
end-to-end packet transfer \cite{ross2012computer}. The transmission
of $X^{N}$ over such link may use short packets, each one encoding
fragments of size $L$. Then, similarly to the standard distributed
compression problem, i.e., the Slepian-Wolf problem \cite{slepian1973noiseless},
the reconstruction of the sequence from the fragments can be aided
by a side-information sequence $Y^{N}$. Third, as we discuss below,
the problem is related to the identification of unordered entities
marked by barcodes from noisy fragments of those barcodes \cite{tandon2019bee}.

In this paper, we assume that $X^{N}$ is drawn from a memoryless
source, and that $Y^{N}$ is obtained by passing $X^{N}$ in a discrete
memoryless channel $P_{Y|X}$. Furthermore, we assume that the length
of each fragment scales logarithmically with the number of fragments,
and specifically set $L=\beta\log M$ for some length-scaling parameter
$\beta$. For this scaling, the problem described above is closely
related, and perhaps \emph{prima facie} equivalent, to the \emph{bee
identification }(BI) problem, recently introduced in \cite{tandon2019bee}
and further studied in \cite{tandon2020bee,tamir2021error,chrisnata2022bee,kiah2022efficient}.
In the BI problem, one assumes that the fragments of $Y^{N}$ (obtained
similarly to the fragments of $X^{N}$, as non-overlapping, consecutive
segments of length $L$) are each a barcode used for identification
of some objects, via the fragments of $X^{N}$, which are noisy unordered
observations of the barcodes. A codebook for this problem comprises
of the $M$ fragments of $Y^{N}$, where $X^{N}$ is drawn in a memoryless
fashion according to the reverse channel $P_{X|Y}$. A plausible method
to generate this codebook is via random coding, and specifically,
by drawing the $N=LM$ symbols of the fragments in an IID fashion.
In this random coding regime, the \emph{average }error probability
over the random ensemble of codebooks is similar to the reconstruction
error in the fragments reordering problem we consider, with the inconsequential
difference that the channel for the BI problem is the reverse channel
$P_{X|Y}$, rather than $P_{Y|X}$. 

Nonetheless, there are two subtle, yet significant, differences between
the ordering problem using reference sequences considered here and
the BI problem. First, in the ordering problem, one is not necessarily
interested in recovering the exact permutation of the fragments, but
rather just the correctly reordered sequence. Second, the source sequence
and the reference sequence are random, and there is no design freedom
to optimally choose the source fragments. By contrast, in the BI problem,
only a single optimal codebook is sought. As shown in \cite{tandon2019bee,tamir2021error},
improved bounds can be obtained by considering the average error of
the \emph{typical random code} \cite{barg2002random,merhav2018error},
or via expurgation techniques \cite{tamir2021error}. As said, in
the ordering problem considered here, this is impossible, and thus
it is the random coding analysis that is of interest. In fact, these
two matters are interrelated, as exemplified by the following extreme
case: In the event that all fragments of $X^{N}$ are equal, there
is no ambiguity in the reconstruction of the sequence, and \emph{zero
reconstruction error} is obtained. By contrast, as a codebook for
the BI problem, this has the \emph{maximal} possible error probability.
Generalizing upon this observation, the difference between the ordering
problem and the BI problem is most pronounced whenever there are \emph{repeated
fragments }in the sequence. This typically happens when the fragments
are relatively short (small $\beta$), and the entropy of the source
probability mass function (PMF) $P_{X}$ is low. 

When considering repeated fragments in a sequence, it is also expected
that \emph{similar} fragments will also be observed, and in such a
scenario, it is unreasonable to expect a perfect reconstruction. Therefore,
we consider in this paper a relaxed notion of \emph{imperfect }reconstruction,
comprised of two elements. First, we assume that an additive distortion
measure is given, and consider a fragment to be successfully reconstructed
if its distortion with the source fragment in that location is below
a prescribed distortion level $\delta\in\mathbb{R}_{+}$. Second,
we consider the reconstruction to be successful if at most a fraction
$\xi\in[0,1]$ of the fragments were successfully reconstructed (i.e.,
their distortion level is less than $\delta$). We then may analyze
the failure probability of the reconstruction algorithm for a pair
$(\delta,\xi)$, or, the trade-off between $\delta$ and $\xi$ .
The relaxed definition of failure probability for $\xi>0$ was also
proposed in the conclusion part of \cite{tandon2019bee} for the BI
problem, as well as in \cite{tamir2021error} (although with $\xi M$
being replaced by a constant that does not scale with $M$). 

As might be noted, and owing to the above described differences, we
describe our setting with a different terminology compared to the
way it is formulated in the BI problem (though they are equivalent).
In the BI problem, the fragment length is considered the decoding
blocklength, and is expected to be large. The number of bees is then
exponential in that blocklength, that is, $M=e^{\frac{L}{\beta}}$.\footnote{In the notation in \cite{tandon2019bee}, $L\leftrightarrow n$ and
$M\leftrightarrow m$.} Therefore, $1/\beta$ may be perceived as the \emph{rate} of identification,
and reliable identification of the bees is shown to be possible in
\cite{tandon2019bee,tamir2021error} as long as $\frac{1}{\beta}$
is less than a maximal possible rate, which may be considered the
\emph{capacity} of the BI problem. Here, we opt to equivalently refer
to the fragment length as a logarithmic function of the sequence length,
as common in various other fragmented sequences problems, such as
DNA storage \cite{shomorony2021dna,weinberger2022dna,weinberger2022Error}.
Accordingly, the success of reconstruction will be (equivalently)
stated in terms of \emph{lower} bounds on the fragment length scaling
$\beta$. 

\subsection{Results Overview}

We assume throughout that the optimal maximum likelihood (ML) decoder
is used for reconstruction, that is, \emph{joint decoding} of all
fragments in the parlance of \cite{tandon2019bee}. 

\subsubsection{The No-Repeating-Fragments Regime with Zero Distortion}

First, we consider the regime in which no repeated fragments are expected,
and assume zero distortion $\delta=0$, though we allow both perfect
and imperfect reconstruction ($\xi=0$ and $\xi>0$, respectively).
For $\xi=0$, this revisits the setting of random coding analysis
for joint decoding in the BI problem \cite{tandon2019bee}. We show
the following in Theorem \ref{thm: Critical beta for no repeat read regime}:
As long as $\beta>\frac{1}{\psi_{2}(P_{XY})}$, then the reconstruction
algorithm succeeds with high probability, where the threshold $\psi_{2}(P_{XY})$
is explicitly defined in (\ref{eq: rate function for a Chernoff upper bound order 2})
as a convex optimization problem over joint PMFs $Q_{X_{1}X_{2}}$,
that is, over $|{\cal X}|^{2}-1$ free variables. For example, when
$X$ is a uniform Bernoulli random variable (RV), and $P_{Y|X}$ is
a binary symmetric channel (BSC) with crossover probability $\alpha$
then $\psi_{2}(P_{XY})=\frac{1}{2}[\log2-\log(1+4\alpha(1-\alpha))]$.
Specifically, if $\xi=0$ then failure occurs with probability at
most $O(M^{2[1-\beta\psi_{2}(P_{XY})]})$, that is, a polynomial decay
in $M$. If $\xi>0$ then the failure probability occurs with probability
at most $e^{-\xi M\log M\cdot[\beta\psi_{2}(P_{XY})-1]}$, that is,
exponential decay with respect to (w.r.t.) $M\log M$. We then establish
in lower bounds in Theorem \ref{thm: tightness of no-repeat}, which
make, mild, unavoidable assumptions. For $\xi=0$, we show that the
failure probability rate is in fact tight for $\xi=0$, and lower
bounded as $e^{-\xi M\log M\cdot\beta\psi_{2}(P_{XY})}$ for $\xi>0$
(thus, there is a gap of $\xi$ in the exponent w.r.t. $M\log M$). 

In the $\xi=0$ setting, the improvement of Theorem \ref{thm: Critical beta for no repeat read regime}
over \cite{tandon2019bee} is twofold. First, \cite{tandon2019bee}
only considered a symmetric (uniform) binary source with a BSC, and
its analysis heavily utilizes the symmetry properties of this distribution.
We obtain this result for a \emph{general }source\emph{ $P_{XY}$}.
In addition, our bound is tighter than the one obtained for the binary
symmetric setting considered in \cite{tandon2019bee}. Specifically,
the dependence on $\psi_{2}(P_{XY})$ is related to the error probability
of transpositions, i.e., cycles of length $2$ (hence the subscript
$2$), and as we show, this is the dominant error event. The analysis
of \cite{tandon2019bee} only showed that the dominant error event
may be either a transposition or a cycle of length $3$. In terms
of proof techniques, as in previous papers, we condition on the source
vector, use a union bound of the pairwise error of all possible permutations,
and upper bound the pairwise error using the Bhattacharyya bound.
As previous analysis also showed, the average of this upper bound
over $X^{N}$ should be computed for permutations which are \emph{cycles.}
Such a cycle may have any length from $\{2,3,...,M-1,M\}$, and we
proved that transpositions (cycles of length $2$) dominate the error
probability. The key new ingredient is to evaluate this expectation
via the \emph{Donsker-Vardhan} variational formula \cite{donsker1983asymptotic}
(e.g., \cite[Corollary 4.15]{boucheron2013concentration} ). This
method is preferred over perhaps the more straightforward \emph{method
of types} \cite{csiszar1998method} \cite[Sec. 2.1]{csiszar2011information},
since the error term in the latter blows up when the cycle length
is on the order of $M$. Interestingly, the argument used to show
that transpositions dominate cycles of length $3$ and the analogous
argument for longer cycles are different. The argument for length-$3$
cycles is direct, and is based on special symmetry properties along
with Han's inequality \cite{han1978nonnegative}. The argument for
cycles of lengths $4$ and larger is based on a \emph{relaxation}
of the function $\psi_{K}(P_{XY})$, characterizing the Bhattacharyya
error bound for length-$K$ cycles, which, intuitively speaking, \emph{breaks}
the cycle at its end point. 

\subsubsection{The Repeating-Fragments Regime with Positive Distortion}

In this regime, we essentially consider both positive fragment-failure
rate $\xi>0$, and positive distortion $\delta>0$. As in the previous
regime and in previous works, the analysis of the reconstruction failure
probability is based on a union bound over all possible permutations
of the $M$ fragments. In the worst case, in which all fragments are
unique, this is a union bound over $M!=e^{M\log M+O(\log M)}$ fragments.
However, in the repeating fragments regime, multiple permutations
are in fact equivalent, in the sense that they lead to the same reconstructed
sequence (e.g., if all fragments are equal except for one, then there
are just $M$ different possible reconstruction vectors). The number
of possible distinct permutations is determined by the histogram of
the $|{\cal X}|^{L}$ possible fragments. Clearly, repeated fragments
are more prone to occur when fragments are short (small $\beta$),
or when the entropy of the source $H(P_{X})$ is low. The main technical
contribution in this regime, and the key ingredient in the analysis,
is to show that the possible number of reconstruction vectors is tightly
concentrated around $e^{\beta H(P_{X})\cdot M\log M+o(M\log M)}$
(see Prop. \ref{prop: number of possible reconstructions}), with
high probability of $1-o(1)$. Thus, if $H(P_{X})\beta<1$ then the
effective number of permutations is smaller than the maximal value
of order $e^{M\log M}$. In turn, the proof of this property is based
on two main ingredients. First, while the histogram vector of the
fragments is distributed as a \emph{multinomial} and thus has dependent
entries, probabilities defined on events of this random vector are
dominated by a Poisson distribution with independent entries (an effect
known as \emph{Poissonization}, see Fact \ref{fact: Poissonization of the multinomial distribution}
and Lemma \ref{lem: Poissonization of events}). The logarithm of
the number of possible reconstruction vectors is then upper bounded
by the entropy of the histogram vector, which using the Poissonization
effect, is the sum of independent terms of the form $-\sum_{i\in[{\cal X}^{L}]}G_{i}\log G_{i}$,
where $G_{i}$ follows a Poisson distribution. For $\beta\in(0,2)$,
the proof then uses a concentration inequality on Lipschitz functions
of Poisson RVs by Bobkov and Ledoux \cite{bobkov1998modified,kontoyiannis2006measure},
however in a modified way, since $t\to-t\log t$ is, strictly speaking,
not a Lipschitz function. For $\beta>2$ a standard Bernstein inequality
is used. The analysis of the failure probability bound then follows
a different path compared to previous works. Rather then fixing a
permutation and analyzing the probability over a random choice of
$X^{N}$, we upper bound the probability for a fixed, typical, $X^{N}$,
in the sense that the number of its possible reconstructions is $e^{\beta H(P_{X})\cdot M\log M}$.
Per the analysis above, it holds that a-typical $X^{N}$occur with
probability at most $o(1)$. Now, if we let $d_{P_{Y|X}}^{*}(\delta)$
be the minimal Bhattacharyya distance for fragments of distortion
larger than $\delta$, it is easily shown that the failure reconstruction
for such typical $X^{N}$ decays as $e^{-\Theta(M\log M)}$ as long
as $\xi>H(P_{X})/d_{P_{Y|X}}^{*}(\delta)$. This leads to a trade-off
between $\delta$ and $\xi$ in the repeating-fragments regime $\beta<1/H(P_{X})$,
which is the main result in this regime, stated in Theorem \ref{thm: Critical beta for repeat read regime}. 

\subsection{Additional Related Work}

An information-theoretic study of sequence reconstruction from short
fragments taken at \emph{random} locations was initiated in \cite{motahari2013information},
and its reference-based counterpart was considered in \cite{mohajer2013reference}.
The analysis in those papers is motivated by DNA sequencing, and thus
assumes a uniform source over the DNA alphabet of size $4$, and a
quaternary symmetric channel $P_{Y|X}$, with the goal of detecting
\emph{single nucleotide polymorphisms} (SNPs). The performance metric
is the average misdetection and false-alarm defined therein (essentially
equivalent to the total number of failed fragments, $\xi M$ in our
notation). In \cite{cassuto2020efficient}, the problem of compressing
a non-probabilistic source was considered when the encoder has a possible
list of reference vectors. In \cite{gershon2021efficient,gershon2022genomic},
compression methods were proposed and analyzed for the setting in
which fragments are compressed at the encoder side and are reconstructed
at the decoder side using a reference sequence. The ordering problem
is also tightly related to the DNA storage sampling-shuffling channel
\cite{shomorony2021dna}, in which short unordered fragments store
the information, or, more generally, to \emph{permutation channels}
\cite{kovavcevic2018codes,makur2020coding,sima2021coding}, in which
the output sequence is a permuted and possibly noisy version of the
input sequence.

\subsection{Outline }

The outline of the rest of the paper is as follows. In Sec. \ref{section:problem formulation}
we formulate the problem, in Sec. \ref{sec:Main-Results} we state
our main results, and in Sec. \ref{sec:Conclusion-and-Future} we
conclude the paper. Proofs are relegated to the appendices. 

\section{Problem Formulation \label{section:problem formulation}}

\paragraph*{Notation conventions}

Let $Q_{X}$ be a PMF over a finite alphabet ${\cal X}$. For $j>i$,
the sequence comprised of the components between $i$ and $j$ is
denoted by $X_{i}^{j}:=(X_{i},X_{i+1},\ldots,X_{j})$ and is shorthanded
as $X^{N}\equiv X_{1}^{N}$ for $i=1$. Let ${\cal P}_{L}({\cal X})$
denote the set of all types (empirical distributions) of length $L$,
and let ${\cal P}({\cal X})$ be the set of all PMFs on ${\cal X}$
(i.e., the $(|{\cal X}|-1)$-dimensional probability simplex). The
type class \cite[Ch. 2]{csiszar2011information} of a type $Q_{X}\in{\cal P}_{L}({\cal X})$
is denoted by $T_{L}(Q_{X})$, that is, the set of all empirical PMFs
for length $L$ vectors over ${\cal X}$. The R\'{e}nyi entropy of
order $\alpha\geq0$, $\alpha\neq1$ is denoted by 
\[
H_{\alpha}(Q_{X}):=\frac{1}{1-\alpha}\log\left(\sum_{x\in{\cal X}}Q_{X}^{\alpha}(x)\right),
\]
and the Shannon entropy is denoted by $H(Q_{X})\equiv H_{1}(Q_{X}):=\lim_{\alpha\downarrow1}H_{\alpha}(Q_{X})=-\sum_{x\in{\cal X}}Q_{X}(x)\log Q_{X}(x)$.
Specifically, $H_{2}(Q_{X})=-\log\sum[Q_{X}(x)]^{2}$ is the \emph{collision
entropy}. The binary entropy function is denoted by $h_{\text{bin}}(t):=-t\log t-(1-t)\log(1-t)$
for $t\in(0,1)$ and $h_{\text{bin}}(0)=h_{\text{bin}}(1)=0$. For
a pair of conditional PMFs $Q_{Y|X}$ and $P_{Y|X}$ and a PMF $P_{X}$,
the conditional Kullback-Leibler (KL) divergence is denoted by $D_{\text{KL}}(Q_{Y|X}\mid\mid P_{Y|X}\mid P_{X})$.
The conditioning on $P_{X}$ is removed when $Y$ is independent of
$X$ under both $Q_{Y|X}$ and $P_{Y|X}$. The binary KL divergence
function is denoted by $d_{\text{bin}}(p,q):=p\log\frac{p}{q}+(1-p)\log\frac{1-p}{1-q}$
for $p,q\in(0,1)$, $d_{\text{bin}}(0,0)=d_{\text{bin}}(1,1)=0$,
and $d_{\text{bin}}(1,0)=d_{\text{bin}}(0,1)=\infty$. The total variation
distance ($\ell_{1}$ distance) between a pair of PMFs over a countable
alphabet is denoted by $d_{\text{TV}}(P,Q):=\sum_{y\in{\cal Y}}|P(y)-Q(y)|$.
The complement of an event ${\cal A}$ is denoted by ${\cal A}^{c}$.
For an integer $M$, $[M]:=\{1,\ldots,M\}$. The maximum (resp. minimum)
between $a$ and $b\in\mathbb{R}$ is denoted by $a\vee b$ (resp.
$a\wedge b$). The maximum between $t\in\mathbb{R}$ and $0$ is denoted
by $(t)_{+}:=t\vee0$.

Let $(X^{N},Y^{N})\sim P_{XY}^{\otimes N}$ be a pair of sequences
of length $N$, drawn IID from $P_{XY}$, over the finite Cartesian
product alphabet ${\cal X}\times{\cal Y}$. The PMF $P_{X}$ is assumed
without loss of generality (WLOG) to be fully supported on ${\cal X}$.
Let $L$ denote a fragment length. For simplicity of notation we assume
that $M:=N/L$ is integer, and ignore in what follows any integer
constraints on asymptotically large numbers, as they are inconsequential
to the results. The sequence $X^{N}$ is partitioned into $M$ equal-length
and non-overlapping fragments denoted by $\boldsymbol{X}(i):=X_{(i-1)L+1}^{iL}$.
A reconstruction algorithm observes the \emph{multiset} of fragments
$\{\boldsymbol{X}(i)\}_{i\in[M]}$ and the reference sequence $Y^{N}$,
and is required to output the original ordered sequence $X^{N}$.
Let $S_{M}$ denote the symmetric group of order $M$, i.e., the group
of all bijections from $[M]$ to itself. A permuted sequence of fragments
is denoted by 
\[
\pi[X^{N}]:=\left(\boldsymbol{X}(\pi(1)),\boldsymbol{X}(\pi(2)),\ldots,\boldsymbol{X}(\pi(M))\right),
\]
and 
\[
{\cal A}_{L}(X^{N}):=\left\{ \pi[X^{N}]\right\} _{\pi\in S_{M}}
\]
is then the \emph{set} of all possible reconstructed sequences from
fragments of $X^{N}$ of length $L$. In essence, conditioned on $X^{N}$,
the reconstruction problem is a multiple hypothesis testing problem
between a \emph{random }number of $|{\cal A}_{L}(X^{N})|$ hypotheses.
An ML reconstruction algorithm chooses an $\hat{X}^{N}\in{\cal A}_{L}(X^{N})$
that satisfies
\[
\hat{X}^{N}=\argmax_{\tilde{X}^{N}\in{\cal A}_{L}(X^{N})}\P\left[Y^{N}\,\middle\vert\,\tilde{X}^{N}\right],
\]
or equivalently, a proper permutation (ordering) of the fragments
$\{\boldsymbol{X}(i)\}_{i\in[M]}$. The ML reconstruction can be cast
as a max-weight matching problem, and thus can be computed in $O(M^{3})$
time \cite{edmonds1972theoretical}, or via message passing algorithms
\cite{cheng2006iterative}. The fragments of $\hat{X}^{N}$ are similarly
denoted by $\hat{\boldsymbol{X}}(i)=\hat{X}_{(i-1)L+1}^{iL}$, and
the fragments of $Y^{N}$ by $\boldsymbol{Y}(i)=Y_{(i-1)L+1}^{iL}$. 

Let $\Delta:{\cal X}\times{\cal X}\to\mathbb{R}_{+}$ be a distortion
measure. With a slight abuse of notation, the distortion measure is
additively extended to length-$L$ fragments $\tilde{\boldsymbol{X}},\overline{\boldsymbol{X}}\in{\cal X}^{L}$
as 
\[
\Delta(\tilde{\boldsymbol{X}},\overline{\boldsymbol{X}})=\frac{1}{L}\sum_{j\in[L]}\Delta(\tilde{X}_{j},\overline{X}_{j}).
\]
Given a desired distortion level $\delta>0$, $\hat{\boldsymbol{X}}(i)$
is said to fail to reconstruct $\boldsymbol{X}(i)$ if $\Delta(\boldsymbol{X}(i),\hat{\boldsymbol{X}}(i))\geq\delta$.
Let 
\[
\Xi_{\delta}(X^{N},\hat{X}^{N}):=\frac{1}{M}\sum_{i\in[M]}\I\{\Delta(\boldsymbol{X}(i),\hat{\boldsymbol{X}}(i))\geq\delta\}
\]
be the relative number of fragments that failed to be properly reconstructed
at distortion level $\delta$. The reconstruction failure probability
at distortion level $\delta\geq0$ and failure level $\xi\in[0,1)$
is then 
\[
\mathsf{FP}(\delta,\xi):=\P\left[\Xi_{\delta}(X^{N},\hat{X}^{N})\geq\xi\right].
\]
Our goal is to establish conditions under which $\mathsf{FP}(\delta,\xi)$
asymptotically vanishes, as $M\to\infty$. We assume that the length
of the fragments scales logarithmically with the number of fragments
$M$, and the scaling is determined by a \emph{fragment length parameter}
$\beta>0$ as
\[
L=\beta\cdot\log M.
\]
Note that it holds for this parametrization that $|{\cal X}|^{L}=M^{\beta}$
and $M=e^{\frac{L}{\beta}}$.

In what follows, the probability of a reconstruction failure will
be bounded using the \emph{Bhattacharyya distance }and more generally,
using the \emph{Chernoff distance}. For a pair of symbols $\overline{x},\tilde{x}\in{\cal X}$,
a transition probability kernel $P_{Y|X}$, and a parameter $s\in[0,1]$,
the Chernoff distance is denoted by 
\begin{equation}
d_{P_{Y|X},s}(\overline{x},\tilde{x}):=-\log\sum_{y\in{\cal Y}}P_{Y|X}^{s}(y\mid\overline{x})\cdot P_{Y|X}^{1-s}(y\mid\tilde{x}).\label{eq: Chernoff distance def}
\end{equation}
For brevity, the dependence of the Chernoff distance on $P_{Y|X}$
will often be suppressed henceforth. Moreover, in most of this paper,
this distance will be used for $s=1/2$. In this case $d_{P_{Y|X},1/2}(\overline{x},\tilde{x})$
is symmetric, it will be referred to as the Bhattacharyya distance,
and $s$ will be omitted from the notation. The Chernoff distance
for a pair of sequences $\overline{\boldsymbol{x}},\tilde{\boldsymbol{x}}\in{\cal X}^{L}$
is additively defined by $d_{s}(\overline{\boldsymbol{x}},\tilde{\boldsymbol{x}}):=\sum_{i\in[L]}d_{s}(\overline{x}_{i},\tilde{x}_{i})$.
This additive distance only depends on the joint type of $(\overline{\boldsymbol{x}},\tilde{\boldsymbol{x}})$.
Accordingly, for a given joint type $Q_{\overline{X}\tilde{X}}\in{\cal P}_{L}({\cal X}^{2})$
for some $L\in\mathbb{N}$, we denote (with a slight abuse of notation)
$d_{s}(Q_{\overline{X}\tilde{X}}):=\frac{1}{L}d_{s}(\overline{\boldsymbol{x}},\tilde{\boldsymbol{x}})$
where $(\overline{\boldsymbol{x}},\tilde{\boldsymbol{x}})\in{\cal T}_{L}(Q_{\overline{X}\tilde{X}})$
is arbitrary. The definition can then be continuously extended to
any joint PMF $Q_{\overline{X}\tilde{X}}$ in the interior of ${\cal P}({\cal X}^{2})$.
Similarly, the distortion $\Delta(\overline{\boldsymbol{x}},\tilde{\boldsymbol{x}})$
between $\overline{\boldsymbol{x}}$ and $\tilde{\boldsymbol{x}}$
only depends on their joint type $Q_{\overline{X}\tilde{X}}$, and
so we also denote it by $\Delta(Q_{\overline{X}\tilde{X}})$. The
definition is then continuously extended to any $Q_{\overline{X}\tilde{X}}$
in the interior of ${\cal P}({\cal X}^{2})$. 

\section{Main Results \label{sec:Main-Results}}

We next describe our results for the no-repeating fragments regime
(Sec. \ref{subsec:The-no-repeating-reads}), and then for the repeating-fragments
regime (Sec. \ref{subsec:The-repeating-reads-regime}).

\subsection{The No-Repeating-Fragments Regime with Zero Distortion \label{subsec:The-no-repeating-reads}}

In this section, we address the regime in which typically all fragments
of $X^{N}$ are unique, and no distortion is allowed $\delta=0$.
We thus abbreviate to reconstruction failure probability to $\mathsf{FP}(\xi)$.
Let 

\begin{equation}
\psi_{2}(P_{XY}):=\min_{Q_{X_{1}X_{2}}\in{\cal P}({\cal X}^{2})}\frac{1}{2}D_{\text{KL}}\left(Q_{X_{1}X_{2}}\mid\mid P_{X}^{\otimes2}\right)+d_{P_{Y|X}}(Q_{X_{1}X_{2}}).\label{eq: rate function for a Chernoff upper bound order 2}
\end{equation}
Essentially, $\psi_{2}(P_{XY})$ is the rate function for a transposition
reconstruction error. Intuitively, $\psi_{2}(P_{XY})$ can be thought
of as follows. For a given pair of fragment sequences $\boldsymbol{x}(1)$
and $\boldsymbol{x}(2)$ that have a joint type $Q_{X_{1}X_{2}}$,
the term $d_{P_{Y|X}}(Q_{X_{1}X_{2}})$ captures how hard it is to
confuse them after observing them through the channel $P_{Y|X}$,
and the term $D_{\text{KL}}(Q_{X_{1}X_{2}}\mid\mid P_{X}^{\otimes2})$
captures how unlikely it is for us to see fragments $\boldsymbol{X}(1)=\boldsymbol{x}(1)$
and $\boldsymbol{X}(2)=\boldsymbol{x}(2)$, when the two fragments
are generated IID according to $P_{X}$. Minimizing over $Q_{X_{1}X_{2}}$
corresponds to finding the worst-case pair of fragments type, which
are most likely to produce a reconstruction error. 

The minimization problem in (\ref{eq: rate function for a Chernoff upper bound order 2})
can be easily solved by using Jensen's inequality to obtain a lower
bound on the minimized argument and show that it is achievable. The
short derivation appears in Appendix \ref{sec: proofs no repeating},
and the result is
\begin{equation}
\psi_{2}(P_{XY}):=-\frac{1}{2}\log\left[\sum_{x_{1},x_{2}\in{\cal X}^{2}}P_{X}(x_{1})P_{X}(x_{2})\cdot e^{-2d_{P_{Y|X}}(x_{1},x_{2})}\right].\label{eq: rate function for a Chernoff upper bound order 2 explicit}
\end{equation}
Furthermore, since $d_{P_{Y|X}}(x,x)=0$, an alternative expression
is 
\begin{equation}
\psi_{2}(P_{XY}):=-\frac{1}{2}\log\left[e^{-H_{2}(P_{X})}+\sum_{x_{1},x_{2}\in{\cal X}^{2}\colon x_{1}\neq x_{2}}P_{X}(x_{1})P_{X}(x_{2})\cdot e^{-2d_{P_{Y|X}}(x_{1},x_{2})}\right],\label{eq: rate function for a Chernoff upper bound order 2 explicit alternative}
\end{equation}
which shows that $\psi_{2}(P_{XY})\to\frac{1}{2}H_{2}(P_{X})$ as
the channel $P_{Y|X}$ approaches a clean channel. 

\subsubsection{An Upper Bound on the Reconstruction Error}
\begin{thm}
\label{thm: Critical beta for no repeat read regime}If $\beta>\frac{1}{\psi_{2}(P_{XY})}$
then for $\xi=0$ 
\[
\mathsf{FP}(\xi=0)=O\left(M^{2[1-\beta\psi_{2}(P_{XY})]}\right)
\]
with a constant that depends on $P_{XY}$, and for $\xi>0$
\[
\mathsf{FP}(\xi)=\exp\left[-M\log M\cdot\xi\left(\beta\psi_{2}(P_{XY})-1-O\left(\frac{1}{M}\right)\right)\right].
\]
\end{thm}

\paragraph*{Discussion}

The bound of Theorem \ref{thm: Critical beta for no repeat read regime}
shows a sharp threshold as a function of $\xi$. For perfect reconstruction
($\xi=0$) the failure probability decays polynomially in $M$, whereas
for imperfect reconstruction ($\xi>0$) it decays exponentially with
$M\log M$, which is much faster. The error bound in the $\xi=0$
case is dominated by transposition errors, that is, an almost perfect
reconstruction of the sequence, except for a single pair of fragments
that has exchanged location. The rate function determining the threshold
is given by $\psi_{2}(P_{XY})$, which can be easily computed for
any $P_{XY}$ as a convex optimization problem over ${\cal P}({\cal X}^{2})$,
see (\ref{eq: rate function for a Chernoff upper bound order 2}).
It can also be noted that the symmetry of the Bhattacharyya distance
and the convexity of the KL divergence imply that the optimal solution
$Q_{X_{1}X_{2}}^{*}$ must have equal marginals, i.e., $Q_{X_{1}}^{*}=Q_{X_{2}}^{*}$.
When $\xi>0$, a wrong placement of less than $\xi M$ fragments is
not considered to be a failure, and so transpositions and other permutations
with $M-K$ fixed points, $K$ fixed, do not lead to a failure. For
$\xi>0$, the dominant error event in this bound turns out to be a
set of $\frac{\xi M}{2}$ transpositions. 

\paragraph*{Proof sketch of Theorem \ref{thm: Critical beta for no repeat read regime}}

The proof of Theorem \ref{thm: Critical beta for no repeat read regime}
first addresses a fixed permutation $\pi\in S_{M}$. For any such
$\pi$, the error is essentially a pairwise error event between $X^{N}$
and its permuted version $\tilde{X}^{N}:=\pi[X^{N}]$. This pairwise
error is bounded using the standard Bhattacharyya upper bound (e.g.,
\cite[Sec. 2.3]{viterbi2009principles}), and then averaged over $X^{N}$.
Finally, using the union bound, the reconstruction failure probability
is upper bounded by summing over all possible permutations. 

Since each permutation is a composition of cycles, as in \cite{tandon2019bee},
we analyze the pairwise error probability average upper bound for
cycles. A main ingredient of the proof is the next lemma, which upper
bounds the expected Bhattacharyya upper bound for a cycle of length
$K$. 
\begin{lem}
\label{lem: Bhat bound rate}Let $X_{1}^{K}\sim P_{X}^{\otimes K}$
IID over a finite alphabet ${\cal X}$. Let $\pi\in S_{K}$ be a cycle
of length $K$, and let $\tilde{X}_{j}=X_{\pi(j)}$ for $j\in[K]$.
Let $P_{Y|X}$ be a transition probability kernel. Then, 
\begin{equation}
\E\left[\exp\left(-d_{P_{Y|X}}(X_{1}^{K},\tilde{X}_{1}^{K})\right)\right]\leq e^{-K\cdot\psi_{2}(P_{XY})},\label{eq: upper bound on the expected Bhat rate}
\end{equation}
where $\psi_{2}(P_{XY})$ is defined in (\ref{eq: rate function for a Chernoff upper bound order 2}). 
\end{lem}
The proof of Lemma \ref{lem: Bhat bound rate} is based on first upper
bounding the expected Bhattacharyya upper bound (left-hand side of
(\ref{eq: upper bound on the expected Bhat rate})) using the \emph{Donsker-Vardhan
variational formula} \cite{donsker1983asymptotic}, \cite[Corollary 4.15]{boucheron2013concentration}.
The resulting upper bound is given by $e^{-K\cdot\psi_{K}(P_{XY})}$,
where the rate function $\psi_{K}(P_{XY})$ is a generalized version
of $\psi_{2}(P_{XY})$ for cycles of length $K$, given as a minimization
problem over ${\cal P}({\cal X}^{K})$ (see (\ref{eq: rate function for expected Bhat bound})
in Appendix (\ref{sec: proofs no repeating})). The proof of the lemma
then continues by establishing that transpositions, i.e., cycles of
length $2$, have the minimal rate function, that is, $\psi_{K}(P_{XY})\geq\psi_{2}(P_{XY})$
for all $K\geq2$. The proof of this claim involves two different
arguments. First, the special symmetry of the case $K=3$ is used
to show that $\psi_{3}(P_{XY})\geq\psi_{2}(P_{XY})$. Specifically,
the Bhattacharyya distance for a length-$3$ cycle is given by $d(Q_{X_{1}X_{2}})+d(Q_{X_{2}X_{3}})+d(Q_{X_{1}X_{3}})$,
which is half of the Bhattacharyya distance of $3$ length-$2$ cycles.
Favorably, the third-order KL divergence involved in the optimization
problem of $\psi_{3}(P_{XY})$, to wit, $D_{\text{KL}}(Q_{X_{1}X_{2}X_{3}}\mid\mid P_{X}^{\otimes3})$,
is analogously lower bounded by the KL divergence of the marginal
pairs using \emph{Han's inequality for the KL divergence} \cite[Theorem 4.9]{boucheron2013concentration}\cite{han1978nonnegative}.
For $K\geq4$, such a symmetry does not seem possible to easily exploit.
Instead, we consider a relaxed lower bound $\psi_{K}(P_{XY})\geq\varphi_{K}(P_{XY})$,
where $\varphi_{K}(P_{XY})$ is obtained by a relaxation of the minimization
problem involved in the definition of $\psi_{K}(P_{XY})$, and show
that $\varphi_{K}(P_{XY})\geq\psi_{2}(P_{XY})$ for all $K\geq4$.
The relaxation from $\psi_{K}(P_{XY})$ to $\varphi_{K}(P_{XY})$,
essentially breaks the cycle, by removing the constraint that $\tilde{X}_{1}=X_{K}$.
This enables to show that the minimizer of $\varphi_{K}(P_{XY})$
in ${\cal P}({\cal X}^{K})$ must satisfy a Markov chain condition
$X_{1}-X_{2}-\cdots-X_{K}$, and consequently reduces the problem
from a $K$-dimensional joint PMF in ${\cal P}({\cal X}^{K})$ to
a simple pairwise joint PMF in ${\cal P}({\cal X}^{2})$. This Markov
condition clearly cannot be satisfied with the original cyclic constraint
of $\tilde{X}_{1}=X_{K}$, and this is why the relaxation from $\psi_{K}(P_{XY})$
to $\varphi_{K}(P_{XY})$ is necessary. Substituting the estimate
of Lemma \ref{lem: Bhat bound rate} to the aforementioned union bound
over all permutations, while taking into account the fact that different
cycles of a permutation are independent, directly leads to the upper
bounds in Theorem \ref{thm: Critical beta for no repeat read regime}. 

\paragraph*{A comparison with \cite{tandon2019bee}}

The setting in \cite{tandon2019bee} assumed that $P_{X}$ is a uniform
binary source $P_{X}(X=0)=P_{X}(X=1)=1/2$, and that $P_{Y|X}$ is
a BSC (as well as $\xi=0$, although the results therein most likely
can be extended to $\xi>0$ in a simple way). For this setting, it
was only established that the worst permutation is either a transposition
(length-$2$ cycle) or a length-$3$ cycle. As we show here, it in
fact holds that the worst case is a transposition, and this holds
for a general $P_{XY}$. The proof of this property leads to the improved
bound on the failure probability with polynomial decrease $O(M^{1-\beta\psi_{2}(P_{XY})})$
compared to $O(M^{1-\beta(\psi_{2}(P_{XY})\vee\psi_{3}(P_{XY}))})$
that can be conjectured from \cite{tandon2019bee} for the general
case. A similar effect holds for the $\xi>0$ case. We finally mention
that the ``break of the cycle'' argument that was use here to relax
$\psi_{K}(P_{XY})$ to $\varphi_{K}(P_{XY})$ is inspired from \cite{tandon2019bee},
in which the contribution of the Bhattacharyya distance of the last
pair of fragments $d(X_{K},\tilde{X}_{K})$ was ignored, in order
to obtain tractable bounds. 

\subsubsection{A Lower Bound on the Reconstruction Error}

We next state a lower bound on $\mathsf{FP}(\xi)$:
\begin{thm}
\label{thm: tightness of no-repeat}Assume that $d_{P_{Y|X}}(x_{1},x_{2})<\infty$
and that 
\begin{equation}
\psi_{2}(P_{XY})<\frac{1}{2}H_{2}(P_{X}).\label{eq: Collision entropy condition on psi}
\end{equation}
If $\beta>\frac{1}{\psi_{2}(P_{XY})}$ then it holds for $\xi=0$
that 
\[
\mathsf{FP}(\xi=0)\geq M^{2[1-\beta\psi_{2}(P_{XY})]+o(1)}
\]
and for $\xi>0$ that 
\[
\mathsf{FP}(\xi)\geq\exp\left[-\xi M\log M\cdot[\beta\psi_{2}(P_{XY})+o(1)]\right].
\]
\end{thm}
Theorem \ref{thm: tightness of no-repeat} establishes the tightness
of the upper bound in Theorem \ref{thm: Critical beta for no repeat read regime}
for $\xi=0$, and suffers from a gap of $\xi M\log M$ in the exponent
for $\xi>0$. 

\paragraph*{The origin of the qualifying assumptions}

The condition $d_{P_{Y|X}}(x_{1},x_{2})<\infty$ is technical, and
related to the uniform continuity of $Q_{X_{1}X_{2}}\to d_{P_{Y|X}}(Q_{X_{1}X_{2}})$
over ${\cal P}({\cal X}^{2})$ required to modify a maximum over types
in ${\cal P}_{L}({\cal X}^{2})$ to a maximum over PMFs in the entire
probability simplex ${\cal P}({\cal X}^{2})$. The condition (\ref{eq: Collision entropy condition on psi})
is related to the fact that if $\boldsymbol{X}(1)=\boldsymbol{X}(2)$
has occurred then the probability that the reconstruction algorithm
erroneously transposes $\boldsymbol{X}(1)$ and $\boldsymbol{X}(2)$
is zero, simply because they are identical (this is where the design
goal in the ordering problem setting defers from that of the BI problem).
This is gauged by the second-order R\'{e}nyi entropy, which is related
to the \emph{collision probability} via $\P[\boldsymbol{X}(1)=\boldsymbol{X}(2)]=e^{-H_{2}(P_{X})}$,
and the assumption assures that this probability is negligible compared
to the probability of erroneous reconstruction exchanging $\boldsymbol{X}(1)$
and $\boldsymbol{X}(2)$, whenever they are different. 

\paragraph*{Proof sketch of Theorem \ref{thm: tightness of no-repeat}}

The proof of Theorem \ref{thm: tightness of no-repeat} first considers
the event in which exchanging the order of $\boldsymbol{X}(i_{1})$
and $\boldsymbol{X}(i_{2})$ for some $i_{1},i_{2}\in[M],i_{1}<i_{2}$
is more likely than the correct order (though this does not imply
that the ML reconstruction will actually have a transposition error
in these locations). The probability of this event can be lower bounded
using the technique of Shannon, Gallager and Berlekamp \cite[Corollary to Thm. 5]{shannon1967lower1}.
In turn, this technique is based on Chernoff's bound, and hence, involves
an optimized version over $s\in[0,1]$ of the Chernoff distance, rather
than the Bhattacharyya distance. Nonetheless, it is shown in the proof
that the optimum is obtained for $s=1/2$. For $\xi=0$, the lower
bound on the reconstruction failure then considers a union over all
possible $\binom{M}{2}=\frac{M(M-1)}{2}$ different transpositions.
As is well known, the union bound clipped to $1$ is order-tight for
independent events (or just pairwise independent events). However,
these transpositions are not pairwise independent events, and so it
is not obvious that the union bound is actually tight in this case.
For $\xi=0$, we use de Caen's inequality \cite{de1997lower} to establish
the tightness of the union bound (as was also used in \cite{tandon2019bee}).
For $\xi>0$, we simply lower bound the error via the error occurs
for some (arbitrary) $\xi M/2$ transpositions. In principle, de Caen's
inequality \cite{de1997lower} may be used for the $\xi>0$ setting
too. However, using a seemingly natural extension of the $\xi=0$
does not lead to an improvement over the simpler bound of a single
set of $\xi M/2$ transpositions. It is possible that de Caen's inequality
is not sufficiently tight for this setting (at least in the way we
have attempted to use it), and the tightness in the $\xi>0$ remains
an open question. 
\begin{example}[$\psi_{2}(P_{XY})$ for binary sources]
\label{exa: binary sources no repeat}Consider a binary source ${\cal X}=\{0,1\}$.
The expression in (\ref{eq: rate function for a Chernoff upper bound order 2 explicit alternative})
results 

\[
\psi_{2}(P_{XY})=\frac{1}{2}\left[H_{2}(p)-\log\left(1+\frac{2p(1-p)}{p^{2}+(1-p)^{2}}\cdot\mathsf{BC}^{2}(P_{Y|X})\right)\right]
\]
where $\mathsf{BC}(P_{Y|X}):=e^{-d_{1\leftrightarrow0}}$ with 
\[
d_{1\leftrightarrow0}:=-\log\sum_{y\in{\cal Y}}\sqrt{P_{Y|X}(y\mid0)\cdot P_{Y|X}(y\mid1)}
\]
is the \emph{Bhattacharyya coefficient}. More specifically, assume
that ${\cal Y}=\{0,1\}$, and $P_{Y|X}$ is a BSC with crossover probability
$\alpha\in[0,1/2]$. Then, $\mathsf{BC}(\text{BSC}(\alpha))=\sqrt{4\alpha(1-\alpha)}$
and then
\[
\psi_{2}(P_{XY})=\frac{1}{2}\left[H_{2}(p)-\log\left(1+\frac{8p(1-p)}{p^{2}+(1-p)^{2}}\cdot\alpha(1-\alpha)\right)\right].
\]
It can be seen that as $\alpha\downarrow0$, it holds that $\psi_{2}(P_{XY})\downarrow\frac{1}{2}H_{2}(p)$.
The noiseless case $\alpha=0$ shows the difference between our problem
(sequence recovering) and the BI problem (permutation recovering):
In our problem the actual exponent for $\alpha=0$ is infinite (zero
error, since $X=Y$ with probability $1$), whereas for the BI problem
$\frac{1}{2}H_{2}(p)$. This agrees with the qualifying condition
of Theorem \ref{thm: tightness of no-repeat}, given by $\psi_{2}(P_{XY})\leq\frac{1}{2}H_{2}(p)$. 
\end{example}
\begin{example}[$\psi_{2}(P_{XY})$ for symmetric general sources]
\label{exa: non-binary sources no repeat}Consider $P_{X}$ to be
uniform over ${\cal X}={\cal Y}$, and let the channel $P_{Y|X}$
be symmetric, in the sense that 
\begin{equation}
P_{Y|X}(y\mid x;\alpha):=\begin{cases}
1-\alpha, & y=x\\
\frac{\alpha}{|{\cal Y}|-1} & \text{otherwise}
\end{cases}\label{eq: symmetric channel}
\end{equation}
(this transition kernel generalizes the BSC to larger alphabets).
The computed value of $\psi_{2}(P_{XY})$ for $P_{XY}=P_{X}\otimes P_{Y|X}(y\mid x;\alpha)$
as a function of $\alpha$ appears in Fig. \ref{fig: psi2 for BSC}.
As might be expected, $\psi_{2}(P_{XY})$ increases with $|{\cal X}|$,
and hence the lower bound on $\beta$ decreases. This agree with intuition
since ordering the fragments is easier for larger entropy sources.
\begin{figure}
\centering{}\includegraphics[scale=0.35]{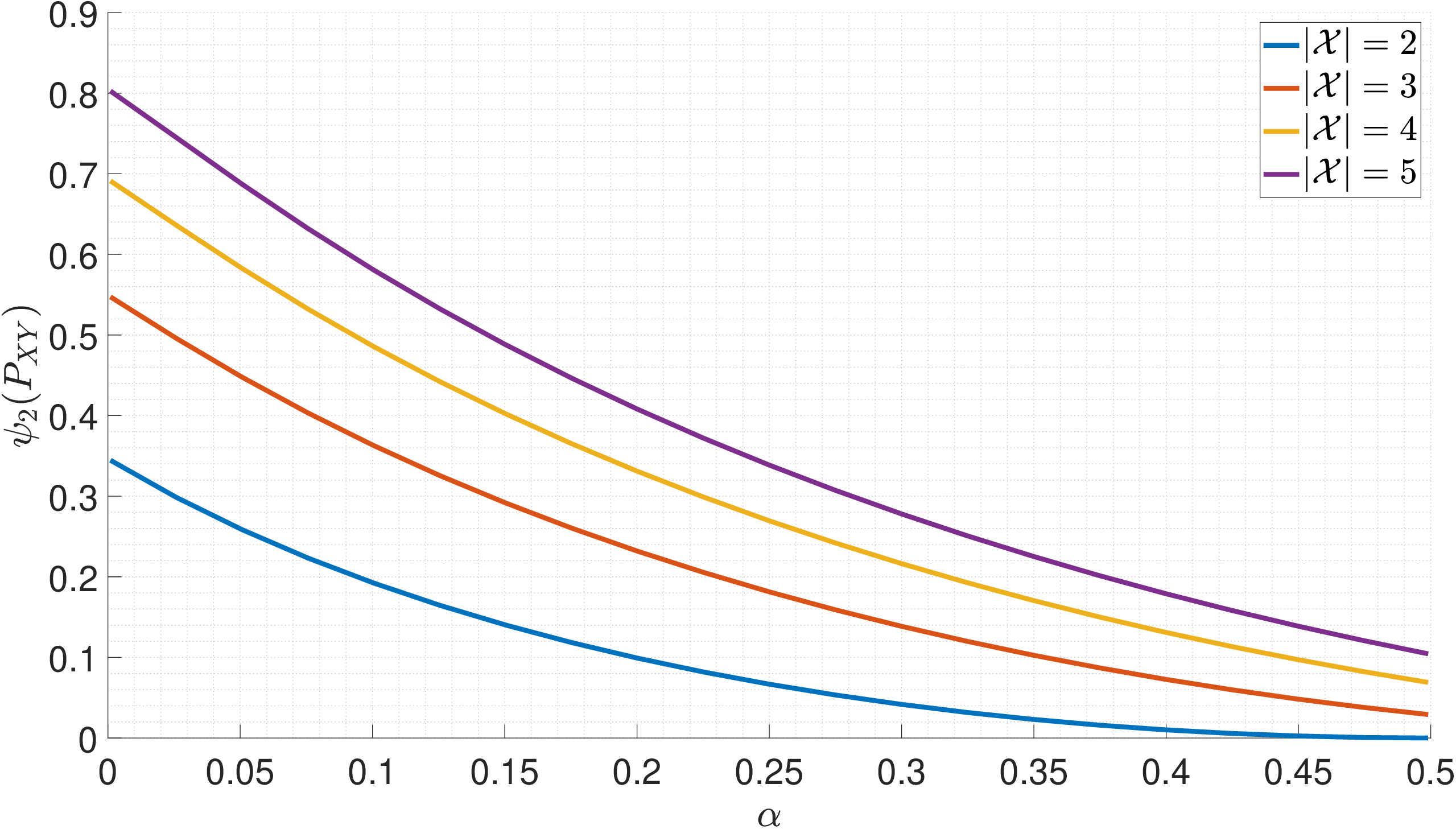}\caption{$\psi_{2}(P_{XY})$ for uniform $P_{X}$ and symmetric channels parameterized
by $\alpha$. \label{fig: psi2 for BSC}}
\end{figure}
\end{example}

\subsection{The Repeating-Fragments Regime with Positive Distortion \label{subsec:The-repeating-reads-regime}}

In this section, we address the regime in which $\beta$ is small,
or the source PMF $P_{X}$ has low entropy. This is the setting in
which the difference between the BI problem and the ordering problem
is most pronounced, since when fragments repeat themselves in the
sequence, reconstruction of the sequence is possible without a reconstruction
of the permutation. In this regime, multiple identical fragments are
typically present in the sequence $X^{N}$. Since in that case fragments
which are similar according to the distortion measure $\Delta$ are
also likely to be abound, we tolerate a positive distortion level.
Intuitively, in this setting, a successful reconstruction is possible,
because if a pair of fragments has distortion larger than the threshold
$\delta$, then it also has large Bhattacharyya distance, and so the
correct order can be identified using the corresponding fragment in
the reference sequence. Concretely, this can be gauged by 
\begin{equation}
d_{P_{Y|X}}^{*}(\delta):=\min_{Q_{X_{1}X_{2}}\in{\cal P}({\cal X}^{2})\colon\Delta(Q_{X_{1}X_{2}})\geq\delta}d_{P_{Y|X}}(Q_{X_{1}X_{2}}),\label{eq: minimal Bhat for a given distortion}
\end{equation}
which is the minimal Bhattacharyya distance possible for any joint
PMF of a pair of fragments whose distortion level is above $\delta$.
Clearly, there is a trade-off between the distortion level $\delta$
and the fraction $\xi$ of failed reconstructed fragments that can
be tolerated -- increasing the distortion level $\delta$ allows
to reduce $\xi$. Our main result in this section characterizes the
trade-off between $\xi$ and $\delta$, which still allows for vanishing
failure probability, as follows: 
\begin{thm}
\label{thm: Critical beta for repeat read regime}Assume that $\beta<\frac{1}{H(P_{X})}$.
Then, if 
\begin{equation}
\xi>\frac{H(P_{X})}{d_{P_{Y|X}}^{*}(\delta)}\label{eq: trade-off between xi and delta}
\end{equation}
then $\mathsf{FP}(\delta,\xi)=o(1).$ 
\end{thm}

\paragraph*{Discussion}

Theorem \ref{thm: Critical beta for repeat read regime} states a
trade-off between $\delta$ and $\xi$ in the repeating-fragments
regime $\beta<1/H(P_{X})$. Interestingly, the minimal possible $\xi$
for a given $\delta$ does not depend on $\beta$ (as long as the
later is sufficiently small). The resulting reconstruction failure
probability then decays to zero, though in an unspecified rate, which
is most likely slower rate compared to the no-repeating fragments
regime, for which the reconstruction failure probability decays as
$e^{-\Theta(\xi M\log M)}$ for $\xi>0$. Evidently, the lower bound
on $\xi$ can be improved by increasing the Bhattacharyya distance,
which can be considered as a measure of the signal strength, or \emph{signal-to-noise
ratio. }Specifically, given any $\xi>0$, the ``quality'' of $P_{Y|X}$
should be such that $d_{P_{Y|X}}^{*}(\delta)\geq H(P_{X})/\xi$. In
other words, any arbitrarily small $\xi>0$ can be compensated by
taking $d_{P_{Y|X}}^{*}(\delta)\to\infty$, that is, making the channel
$P_{Y|X}$ ``cleaner'' (specifically, if $Y=X$ with probability
$1$ then $d_{P_{Y|X}}^{*}(\delta)\uparrow\infty$ for any non-trivial
distortion measure $\Delta$). Theorem \ref{thm: Critical beta for repeat read regime}
states an achievable trade-off between $(\xi,\delta)$ and $\beta$,
and evaluating the tightness of this trade-off and its possible dependence
on $\beta$ is an interesting open problem. 

\paragraph*{Proof of Theorem \ref{thm: Critical beta for repeat read regime}
-- The typical cardinality of the set ${\cal A}_{L}(X^{N})$}

As stated in the problem formulation, the reconstruction problem is
a hypothesis testing problem between a \emph{random} number of $|{\cal A}_{L}(X^{N})|$
hypotheses, or equivalently, all possible \emph{different} reconstructed
sequences. Upper bounds on the error probability in multiple hypothesis
testing typically involve some sort of a union bound over the alternative
hypotheses, and similarly so is our upper bound on the failure probability.
Therefore, a main technical part is to establish a tight upper bound
on the number of alternative hypotheses. If all fragments $\{\boldsymbol{X}(i)\}_{i\in[M]}$
are unique, then the number of possible reconstruction vectors is
$M!=e^{M\log M+O(M)}$. However, if the source PMF $P_{X}$ is such
that some fragments in ${\cal X}^{L}$ are expected to repeat multiple
times, then it is expected that $\log|{\cal A}_{L}(X^{N})|$ will
be significantly smaller than $M\log M+O(M)$. The main ingredient
of the analysis of the reconstruction failure in this regime shows
that $\log|{\cal A}_{L}(X^{N})|\leq\beta H(P_{X})\cdot M\log M$ essentially
holds with probability $1-o(1)$. This cardinality can be much smaller
for low $\beta$ or sources with low entropy. 

To accurately present this bound, let us assume for notational simplicity,
that the $L$th order Cartesian product of ${\cal X}$ is arbitrarily
ordered as ${\cal X}^{L}\equiv\{a_{1}\ldots,a_{M^{\beta}}\}$, where
we recall that $|{\cal X}^{L}|=M^{\beta}$. Then, for any given vector
$x^{N}\in{\cal X}^{N}$ and any $j\in[M^{\beta}]$,
\[
g_{L}(j;x^{N}):=\sum_{i\in[M]}\I\{\boldsymbol{x}(i)=a_{j}\}
\]
is the number of times that the length-$L$ vector $a_{j}\in{\cal X}^{L}$
appears in the fragments of $x^{N}$, and
\[
g_{L}(x^{N}):=\left(g_{L}(1;x^{N}),g_{L}(2;x^{N}),\ldots,g_{L}(M^{\beta};x^{N})\right)\in[M+1]^{M^{\beta}}
\]
is the \emph{histogram} vector of $x^{N}$ for length-$L$ fragments.
It holds that $\sum_{j\in[M^{\beta}]}g_{L}(j;x^{N})=M$. For brevity,
we next denote the random number of appearances of the $j$th letter
of ${\cal X}^{L}$ in the $M$ fragments of $X^{N}$ as $G(j):=g_{L}(j;X^{N})$.
The formal bound is as follows:
\begin{prop}
\label{prop: number of possible reconstructions}Assume that $H(P_{X})>0$.
There exists a constant $c>0$ so that for any $\eta\in(0,1)$, the
log-cardinality of ${\cal A}_{L}(X^{N})$ is concentrated as 
\[
\P\left[\frac{1}{M}\log\left|{\cal A}_{L}(X^{N})\right|\geq L\cdot H(P_{X})+\eta\log M\right]=\begin{cases}
\exp\left[-\Omega(\cdot\eta^{2}M^{1\vee(2-\beta)})\right], & 0<\beta<2\\
\frac{2}{M^{\eta/2}}, & \beta\geq2
\end{cases},
\]
for all $M\geq M_{0}(P_{X},\beta,\eta)$. 
\end{prop}
The proof of Prop. \ref{prop: number of possible reconstructions},
which is fully presented in Appendix \ref{sec: proofs repeating},
is based on the standard entropy bound on the multinomial coefficient,
which then leads to the bound
\[
\frac{1}{M}\log\left|{\cal A}_{L}(X^{N})\right|\leq-\sum_{j\in[M^{\beta}]}\frac{G(j)}{M}\log\frac{G(j)}{M}.
\]
Given the fragments model, the histogram vector $G=(G(1),\ldots,G(M^{\beta}))$
is distributed as a \emph{multinomial} RV, and thus its components
are statistically \emph{dependent}. The upper bound on $\frac{1}{M}\log|{\cal A}_{L}(X^{N})|$
is thus a complicated function of this random vector, and so it is
difficult to directly analyze its random perturbation around its mean.
Nonetheless, as is well known, the probability of an event under the
multinomial distribution can be upper bounded by the probability of
the same event under a properly defined Poisson distribution that
has \emph{independent} components. We thus consider a Poissonized
version $\tilde{G}$ of $G$, and analyze the tail behavior of $f(g)\colon\mathbb{N}_{+}\to\mathbb{R}$
for $f(g):=-\frac{t}{M}\log\frac{t}{M}$. For $\beta\in(0,2)$ we
show using concentration bounds for Lipschitz functions of Poisson
RVs \cite{bobkov1998modified} that $f(g)$ is a sub-gamma random
variable \cite[Ch. 2]{boucheron2013concentration}, and then bound
the concentration of $\sum_{j\in[M^{\beta}]}f(\tilde{G}(j))$ via
Bernstein's inequality. A truncation argument is required since, strictly
speaking, $f(g)$ involved in the upper bound is not Lipschitz continuous
on $\mathbb{N}_{+}$. For $\beta>2$ we use a standard Bernstein's
inequality, after using looser bounding techniques. 
\begin{example}[A symmetric channel and Hamming distortion measure]
\label{exa: binary sources  repeat}Assume that ${\cal X}={\cal Y}$
and that $P_{Y|X}$ is a symmetric channel parameterized by $\alpha$,
as in (\ref{eq: symmetric channel}). In this case, it holds that
\textbf{
\[
d_{P_{Y|X}^{(\alpha)}}(x,\tilde{x})=d_{\alpha}\cdot\I[\overline{x}\neq\tilde{x}]
\]
}where for any $\overline{x},\tilde{x}\in{\cal X}$\textbf{ }with\textbf{
$\overline{x}\neq\tilde{x}$
\begin{align}
d_{\alpha} & :=-\log\sum_{y\in{\cal Y}}\sqrt{P_{Y|X}(y\mid\overline{x})\cdot P_{Y|X}(y\mid\tilde{x})}\\
 & =-\log\Bigg[\sqrt{P_{Y|X}(\tilde{x}\mid\overline{x})\cdot P_{Y|X}(\tilde{x}\mid\tilde{x})}+\sqrt{P_{Y|X}(\overline{x}\mid\overline{x})\cdot P_{Y|X}(\overline{x}\mid\tilde{x})}\nonumber \\
 & \hphantom{=========}+\sum_{y\in{\cal X}\backslash\{\overline{x},\tilde{x}\}}\sqrt{P_{Y|X}(y\mid\overline{x})\cdot P_{Y|X}(y\mid\overline{x})}\Bigg]\\
 & =-\log\left[\sqrt{\frac{\alpha}{|{\cal Y}|-1}\cdot(1-\alpha)}+\sqrt{(1-\alpha)\cdot\frac{\alpha}{|{\cal Y}|-1}}+(|{\cal Y}|-2)\cdot\alpha\right]\\
 & :=-\log\left[\sqrt{\frac{4(1-\alpha)\alpha}{|{\cal Y}|-1}}+\frac{(|{\cal Y}|-2)\cdot\alpha}{|{\cal Y}|-1}\right].
\end{align}
}Further assume that the distortion measure is the Hamming distortion
measure $\Delta(\overline{x},\tilde{x})=\I[\overline{x}\neq\tilde{x}]$.
Thus, $d_{P_{Y|X}^{(\alpha)}}(x,\tilde{x})\propto\Delta(\overline{x},\tilde{x})$
and then it is simple to obtain that $d^{*}(\delta)=\delta\cdot d_{\alpha}$,
and the bound of Theorem \ref{thm: Critical beta for repeat read regime}
results in 
\[
\xi>\frac{H(P_{X})}{\delta\cdot d_{\alpha}}.
\]
The achievable trade-off between $\xi$ and $\delta$ is shown in
Fig. \ref{fig: Minimal required channel level} for $\alpha=0.1$
and $H(P_{X})=0.1\,\text{[nats]}$, for varying alphabet sizes. As
can be seen, the minimal $\xi$ is improving for larger alphabet sizes,
though this improvement has diminishing returns. We finally remark
that computing $d_{P_{Y|X}}^{*}(\delta)$ for general channels is
a simple linear program (\ref{eq: minimal Bhat for a given distortion}),
and thus can be easily computed for any arbitrary $P_{Y|X}$ and distortion
measure $\Delta$.
\begin{figure}
\centering{}\includegraphics[scale=0.35]{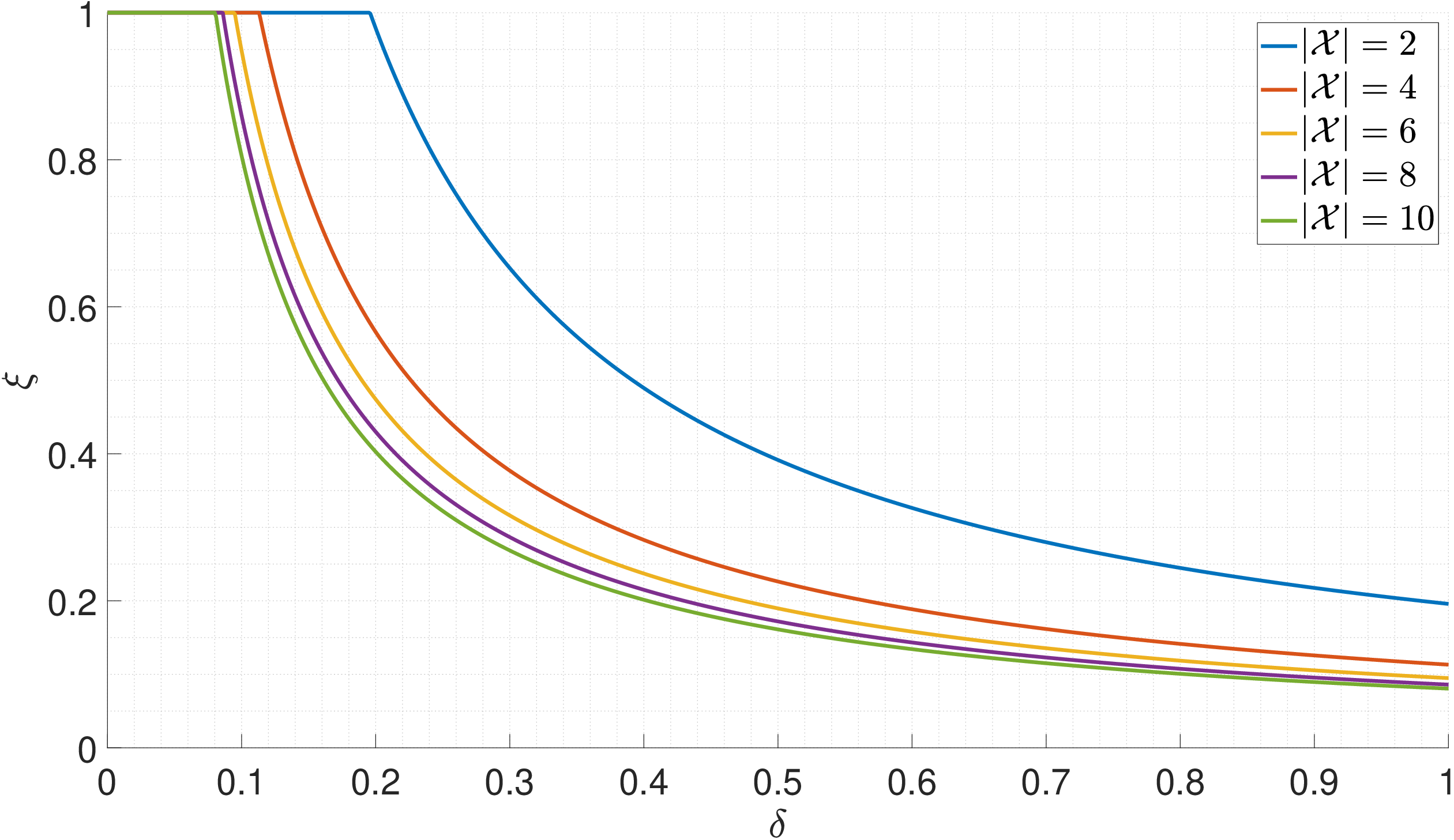}\caption{The trade-off between $\xi$ and $\delta$ for $H(P_{X})=0.1\text{[nats]}$
and symmetric channels $P_{Y|X}^{(\alpha)}$ for $\alpha=0.1$. \label{fig: Minimal required channel level}}
\end{figure}
\end{example}

\section{Conclusion and Future Research \label{sec:Conclusion-and-Future}}

We have considered the problem of ordering the multiset of consecutive
fragments of a sequence, based on a reference sequence. We have considered
a general setting in which each fragment can be reconstructed with
possibly some low distortion level, as well as a constant fraction
of the fragments to be reconstructed with high distortion. First,
we considered the regime in which typically all fragments are unique,
and so focused on zero distortion. For a general joint source $P_{XY}$,
we have derived an upper bound on the fragment length required for
reliable reconstruction, both for perfect and partial reconstruction.
These results tighten and extend previous results derived for the
BI problem in the random coding regime that were restricted to uniform
binary sources $P_{X}$ and symmetric $P_{Y|X}$. In the perfect reconstruction
setting (\textbf{$\xi=0$}) the bound was proved to be tight, whereas
a lower bound was derived for the partial reconstruction ($\xi>0$)
setting. There is a gap between the bounds in the latter setting,
which is an interesting and challenging open problem, mainly since
the using the standard technique based on de Caen's inequality appears
to fail. Second, we considered the regime in which repeating fragments
abound, and showed its relation to the entropy of the source $P_{X}$.
In this regime, it is natural to tolerate a positive distortion $\delta>0$
between the fragments and their reconstruction. We show that as long
as $\beta$ is small enough $(\beta<1/H(P_{X}))$ and the reconstruction
algorithm operates in the repeating-fragments regime, a trade-off
(\ref{eq: trade-off between xi and delta}) is obtained between the
minimal $\xi$ possible for the given $\delta$, which still assures
vanishing failure probability. As said, evaluating the tightness of
the trade-off is an interesting open problem, and specifically, whether
the optimal trade-off depends on $\beta$ or not. Furthermore, it
is of interest to investigate the optimal decay rate of the reconstruction
failure probability, and how it depends on the problem parameters. 

Other avenues for future research include: (i) Reconstruction with
possibly overlapping fragments and high coverage, taken at random
locations (as in \cite{mohajer2013reference}) (ii) Reconstruction
with a compressed version of either the fragments or the reference
sequence (or both), (iii) Reconstruction using fragments that were
obtained from $Y^{N}$ via other channels, e.g., those which include
deletions or insertions (or both), (iv) Reconstruction using multiple
reference fragments, and more. 

\appendices{\numberwithin{equation}{section}}

\section{Proofs for the No-Repeating-Fragments Regime with Zero Distortion
\label{sec: proofs no repeating}}

\paragraph*{Proof of (\ref{eq: rate function for a Chernoff upper bound order 2 explicit})
and (\ref{eq: rate function for a Chernoff upper bound order 2 explicit alternative})}

The argument in the optimization problem (\ref{eq: rate function for a Chernoff upper bound order 2})
defining $\psi_{2}(P_{XY})$ satisfies for any $Q_{X_{1}X_{2}}\in{\cal P}({\cal X}^{2})$
\begin{align}
 & \frac{1}{2}D_{\text{KL}}\left(Q_{X_{1}X_{2}}\mid\mid P_{X}^{\otimes2}\right)+d_{P_{Y|X}}(Q_{X_{1}X_{2}})\nonumber \\
 & =-\frac{1}{2}\sum_{x_{1},x_{2}\in{\cal X}^{2}}Q_{X_{1}X_{2}}(x_{1},x_{2})\left[\log\frac{P_{X}(x_{1})P_{X}(x_{2})}{Q_{X_{1}X_{2}}(x_{1},x_{2})}-2d_{P_{Y|X}}(x_{1},x_{2})\right]\\
 & =-\frac{1}{2}\sum_{x_{1},x_{2}\in{\cal X}^{2}}Q_{X_{1}X_{2}}(x_{1},x_{2})\log\frac{P_{X}(x_{1})P_{X}(x_{2})e^{-2d_{P_{Y|X}}(x_{1},x_{2})}}{Q_{X_{1}X_{2}}(x_{1},x_{2})}\\
 & \trre[\geq,a]-\frac{1}{2}\log\sum_{x_{1},x_{2}\in{\cal X}^{2}}P_{X}(x_{1})P_{X}(x_{2})e^{-2d_{P_{Y|X}}(x_{1},x_{2})},
\end{align}
where $(a)$ follows from Jensen's inequality for the convex function
$t\to-\log t$, and equality is achieved when the averaged arguments
are all equal, that is,
\[
Q_{X_{1}X_{2}}(x_{1},x_{2})=\frac{P_{X}(x_{1})P_{X}(x_{2})e^{-2d_{P_{Y|X}}(x_{1},x_{2})}}{\sum_{x_{1}',x_{2}'\in{\cal X}^{2}}P_{X}(x_{1}')P_{X}(x_{2}')e^{-2d_{P_{Y|X}}(x_{1}',x_{2}')}}.
\]
This proves (\ref{eq: rate function for a Chernoff upper bound order 2 explicit}).
The expression in (\ref{eq: rate function for a Chernoff upper bound order 2 explicit alternative})
follows directly from the definition of the second-order R\'{e}nyi
entropy. 

In order to prove Theorem \ref{thm: Critical beta for no repeat read regime},
we begin by proving Lemma \ref{lem: Bhat bound rate}. 
\begin{IEEEproof}[Proof of Lemma \ref{lem: Bhat bound rate}]
We denote the length-$K$ cycle, in a two-line notation, as 
\[
\pi_{K}^{\circ}:=\left(\begin{array}{cccccc}
1 & 2 & 3 & \cdots & K-1 & K\\
K & 1 & 2 & \cdots & K-2 & K-1
\end{array}\right).
\]
By the variational representation of Donsker-Vardhan \cite{donsker1983asymptotic}
(e.g., \cite[Corollary 4.15]{boucheron2013concentration} ), for
any $Q_{X_{1}X_{2}\ldots X_{K}}\in{\cal P}({\cal X}^{K})$
\[
D_{\text{KL}}(Q_{X_{1}X_{2}\ldots X_{K}}\mid\mid P_{X}^{\otimes K})+\E_{Q_{X_{1}X_{2}\ldots X_{K}}}\left[d(X_{1}^{K},\tilde{X}_{1}^{K})\right]\geq-\log\E_{P_{X}^{\otimes K}}\left[\exp\left(-d(X_{1}^{K},\tilde{X}_{1}^{K})\right)\right].
\]
Minimizing over $Q_{X_{1}X_{2}\ldots X_{K}}$ while using that $P_{X}^{\otimes K}$
has full support and thus $P_{X}^{\otimes K}\gg Q_{X_{1}X_{2}\ldots X_{K}}$
holds for any PMF $Q_{X_{1}X_{2}\ldots X_{K}}$, results 
\[
\E\left[\exp\left(-d(X_{1}^{K},\tilde{X}_{1}^{K})\right)\right]\leq e^{-K\cdot\psi_{K}(P_{XY})}
\]
where $\psi_{K}(P_{XY})$ is given by 
\begin{equation}
\psi_{K}(P_{XY}):=\min_{Q_{X_{1}X_{2}\ldots X_{K}}\in{\cal P}({\cal X}^{K})}\frac{1}{K}D_{\text{KL}}\left(Q_{X_{1}X_{2}\ldots X_{K}}\mid\mid P_{X}^{\otimes K}\right)+\frac{1}{K}\sum_{i\in[K]}d(Q_{X_{i}X_{\pi_{K}^{\circ}(i)}}).\label{eq: rate function for expected Bhat bound}
\end{equation}
We next show that $\psi_{K}(P_{XY})\geq\psi_{2}(P_{XY})$ for all
$K\geq2$. We prove this property separately for $K=3$ and $K\geq4$. 

We prove that $\psi_{3}(P_{XY})\geq\psi_{2}(P_{XY})$ by utilizing
Han's inequality for the KL divergence \cite[Thm. 4.9]{boucheron2013concentration}\cite{han1978nonnegative},
which states that for any probability measure $Q_{Z_{1}Z_{2}\cdots Z_{K}}\in{\cal P}({\cal Z}^{K})$
and a product probability measure $P_{Z_{1}}\otimes P_{Z_{2}}\cdots\otimes P_{Z_{K}}\in{\cal P}({\cal Z}^{K})$
it holds that
\begin{multline}
D_{\text{KL}}(Q_{Z_{1}Z_{2}\cdots Z_{K}}\mid\mid P_{Z_{1}}\otimes P_{Z_{2}}\cdots\otimes P_{Z_{K}})\\
\geq\frac{1}{K-1}\sum_{j\in[K]}D_{\text{KL}}\left(Q_{Z_{1}\cdots Z_{j-1}Z_{j+1}\cdots Z_{K}}\mid\mid P_{Z_{1}}\otimes\cdots P_{Z_{j-1}}\otimes P_{Z_{j+1}}\cdots\otimes P_{Z_{K}}\right),\label{eq: Hans inequality application}
\end{multline}
where $Q_{Z_{1}\cdots Z_{j-1}Z_{j+1}\cdots Z_{K}}$ is understood
as the joint PMF of $Z_{1}^{K}$ marginalized over $Z_{j}$. Indeed,
it then holds that 
\begin{align}
\psi_{3}(P_{XY}) & =\min_{Q_{X_{1}X_{2}X_{3}}}\frac{1}{3}D_{\text{KL}}\left(Q_{X_{1}X_{2}X_{3}}\mid\mid P_{X}^{\otimes3}\right)+\frac{1}{3}\sum_{i=1}^{3}d(Q_{X_{i}X_{\pi_{K}^{\circ}(i)}})\\
 & \trre[\geq,a]\min_{Q_{X_{1}X_{2}X_{3}}}\frac{1}{3}\frac{1}{2}\Bigg[D_{\text{KL}}\left(Q_{X_{1}X_{2}}\mid\mid P_{X}^{\otimes2}\right)+2d(Q_{X_{1}X_{2}})+D_{\text{KL}}\left(Q_{X_{1}X_{3}}\mid\mid P_{X}^{\otimes2}\right)\nonumber \\
 & \hphantom{===}+2d(Q_{X_{1}X_{3}})+D_{\text{KL}}\left(Q_{X_{2}X_{3}}\mid\mid P_{X}^{\otimes2}\right)+2d(Q_{X_{2}X_{3}})\Bigg]\\
 & \geq\frac{1}{3}\frac{1}{2}\Bigg[\min_{Q_{X_{1}X_{2}}}\left\{ D_{\text{KL}}\left(Q_{X_{1}X_{2}}\mid\mid P_{X}^{\otimes2}\right)+2d(Q_{X_{1}X_{2}})\right\} +\nonumber \\
 & \hphantom{===}\min_{Q_{X_{1}X_{3}}}\left\{ D_{\text{KL}}\left(Q_{X_{1}X_{3}}\mid\mid P_{X}^{\otimes2}\right)+2d(Q_{X_{1}X_{3}})\right\} +\min_{Q_{X_{2}X_{3}}}\left\{ D_{\text{KL}}\left(Q_{X_{2}X_{3}}\mid\mid P_{X}^{\otimes2}\right)+2d(Q_{X_{2}X_{3}})\right\} \Bigg]\\
 & =\psi_{2}(P_{XY}),\label{eq: psi_3 is larger than psi_2}
\end{align}
where $(a)$ holds by Han's inequality.

We now turn to prove that $\psi_{K}(P_{XY})\geq\psi_{2}(P_{XY})$
for all $K\geq4$. To this end, consider the minimization problem
involved in the upper bound rate function $\psi_{K}(P_{XY})$, to
wit,
\begin{equation}
\min_{Q_{X_{1}X_{2}\ldots X_{K}}}\frac{1}{K}D_{\text{KL}}\left(Q_{X_{1}X_{2}\ldots X_{K}}\mid\mid P_{X}^{\otimes K}\right)+\frac{1}{K}\sum_{i\in[K]}d(Q_{X_{i}X_{\pi_{K}^{\circ}(i)}}).\label{eq: rate function for expected Bhat bound optimization}
\end{equation}
Now, suppose that $Q_{X_{1}\ldots X_{K}}^{(0)}$ is a solution of
the minimization problem in (\ref{eq: rate function for expected Bhat bound optimization}).
Then, due to the circular symmetry of the objective function of (\ref{eq: rate function for expected Bhat bound optimization}),
\[
Q_{X_{1}\ldots X_{K}}^{(1)}=Q_{X_{\pi_{K}^{\circ}(1)}\ldots X_{\pi_{K}^{\circ}(K)}}^{(0)}
\]
attains the same value for the objective function. We may then recursively
define 
\[
Q_{X_{1}\ldots X_{K}}^{(j)}=Q_{X_{\pi_{K}^{\circ}(1)}\ldots X_{\pi_{K}^{\circ}(K)}}^{(j-1)}
\]
for all $j\in[K-1]\backslash\{1\}$, and similarly, each $Q_{X_{1}\ldots X_{K}}^{(j)}$
also attains the same value for the objective function. Since the
KL divergence is convex and the Bhattacharyya distance is linear in
$Q_{X_{1}X_{2}\ldots X_{K}}$, the objective function in (\ref{eq: rate function for expected Bhat bound})
is convex in $Q_{X_{1}X_{2}\ldots X_{K}}$. Thus, 
\[
\overline{Q}_{X_{1}\ldots X_{K}}=\frac{1}{K-1}\sum_{j=0}^{K-1}Q_{X_{1}\ldots X_{K}}^{(j)}
\]
may only attain a lower value for the objective function. Moreover,
$\overline{Q}_{X_{1}\ldots X_{K}}$ is such that $\overline{Q}_{X_{1}X_{2}}=\overline{Q}_{X_{2}X_{3}}=\cdots=\overline{Q}_{X_{K}X_{1}}$.
Thus, the solution of the minimization problem in (\ref{eq: rate function for expected Bhat bound})
must satisfy that all marginals of consecutive pairs is the same,
let say $Q_{\tilde{X}_{1}\tilde{X}_{2}}\in{\cal P}({\cal X}^{2})$.
Let us define this set of PMFs as 
\[
{\cal Q}_{\text{pairs}}(Q_{\tilde{X}_{1}\tilde{X}_{2}}):=\left\{ Q_{X_{1}X_{2}\cdots X_{K}}\in{\cal P}({\cal X}^{K})\colon Q_{X_{1}X_{2}}=Q_{X_{2}X_{3}}=\cdots Q_{X_{K-1}X_{K}}=Q_{X_{K}X_{1}}=Q_{\tilde{X}_{1}\tilde{X}_{2}}\right\} .
\]
Furthermore, let us define a slightly modified version of this set,
given as 
\[
\hat{{\cal Q}}_{\text{pairs}}(Q_{\tilde{X}_{1}\tilde{X}_{2}}):=\left\{ Q_{X_{1}X_{2}\cdots X_{K}}\in{\cal P}({\cal X}^{K})\colon Q_{X_{1}X_{2}}=Q_{X_{2}X_{3}}=\cdots Q_{X_{K-1}X_{K}}=Q_{\tilde{X}_{1}\tilde{X}_{2}}\right\} ,
\]
where the only difference between ${\cal Q}_{\text{pairs}}(Q_{\tilde{X}_{1}\tilde{X}_{2}})$
and $\hat{{\cal Q}}_{\text{pairs}}(Q_{\tilde{X}_{1}\tilde{X}_{2}})$
is the relaxation of the constraint $Q_{X_{K}X_{1}}=Q_{\tilde{X}_{1}\tilde{X}_{2}}$.
Note that the removal of this constraint effectively ``breaks''
the cycle. Returning to (\ref{eq: rate function for expected Bhat bound}),
we get from this property that 
\begin{align}
\psi_{K}(P_{XY}) & :=\min_{Q_{\tilde{X}_{1}\tilde{X}_{2}}}\left\{ \min_{Q_{X_{1}X_{2}\cdots X_{K}}\colon{\cal Q}_{\text{pairs}}(Q_{\tilde{X}_{1}\tilde{X}_{2}})}\frac{1}{K}D_{\text{KL}}\left(Q_{X_{1}X_{2}\ldots X_{K}}\mid\mid P_{X}^{\otimes K}\right)+d(Q_{\tilde{X}_{1}\tilde{X}_{2}})\right\} \\
 & \geq\min_{Q_{\tilde{X}_{1}\tilde{X}_{2}}}\left\{ \min_{Q_{X_{1}X_{2}\cdots X_{K}}\colon\hat{{\cal Q}}_{\text{pairs}}(Q_{\tilde{X}_{1}\tilde{X}_{2}})}\frac{1}{K}D_{\text{KL}}\left(Q_{X_{1}X_{2}\ldots X_{K}}\mid\mid P_{X}^{\otimes K}\right)+d(Q_{\tilde{X}_{1}\tilde{X}_{2}})\right\} \label{eq: rate function for expected Bhat bound optimization relaxed}\\
 & :=\varphi_{K}(P_{XY}).
\end{align}
This ``cycle-break'' of the set $\hat{{\cal Q}}_{\text{pairs}}(Q_{\tilde{X}_{1}\tilde{X}_{2}})$
and the relaxation of $\psi_{K}(P_{XY})$ to $\varphi_{K}(P_{XY})$
is crucial to establish the following property: The optimal solution
$Q_{X_{1}X_{2}\ldots X_{K}}^{*}$ of $\varphi_{K}(P_{XY})$ must respect
the Markov chain $X_{1}-X_{2}-\cdots X_{K-1}-X_{K}$. Indeed, assume
by contradiction that, under $Q$, this is not the case for $X_{K}$,
that is $Q_{X_{K}|X_{K-1}}\neq Q_{X_{K}|X_{K-1}\cdots X_{1}}$. Then,
\begin{align}
 & D_{\text{KL}}\left(Q_{X_{1}X_{2}\cdots X_{K}}\mid\mid P_{X}^{\otimes K}\right)\nonumber \\
 & =D_{\text{KL}}\left(Q_{X_{1}X_{2}\cdots X_{K-1}}\mid\mid P_{X}^{\otimes(K-1)}\right)+D_{\text{KL}}\left(Q_{X_{K}|X_{K-1}X_{K-2}\cdots X_{1}}\mid\mid P_{X}\mid Q_{X_{1}X_{2}\cdots X_{K-1}}\right)\\
 & \geq D_{\text{KL}}\left(Q_{X_{1}X_{2}\cdots X_{K-1}}\mid\mid P_{X}^{\otimes(K-1)}\right)+D_{\text{KL}}\left(Q_{X_{K}|X_{K-1}}\mid\mid P_{X}\mid Q_{X_{K-1}}\right),\label{eq: a lower bound on the KL divergence if Markov is not satisfied}
\end{align}
where the inequality follows since the convexity of the KL divergence
implies that 
\begin{align}
 & D_{\text{KL}}\left(Q_{X_{K}|X_{K-1}X_{K-2}\cdots X_{1}}\mid\mid P_{X}\mid Q_{X_{1}X_{2}\cdots X_{K-1}}\right)\nonumber \\
 & =\E_{Q_{X_{1}\cdots X_{K-2}|X_{K-1}}}\left[D_{\text{KL}}\left(Q_{X_{K}|X_{K-1}X_{K-2}\cdots X_{1}}(\cdot\mid X_{1}X_{2}\cdots X_{K-1})\mid\mid P_{X}\right)\right]\\
 & \geq D_{\text{KL}}\left(\E_{Q_{X_{1}\cdots X_{K-2}|X_{K-1}}}\left[Q_{X_{K}|X_{K-1}X_{K-2}\cdots X_{1}}\right]\mid\mid P_{X}\mid Q_{X_{1}X_{2}\cdots X_{K-1}}\right)\\
 & \geq D_{\text{KL}}\left(Q_{X_{K}|X_{K-1}}\mid\mid P_{X}\mid Q_{X_{K-1}}\right).
\end{align}
Thus, we can replace any $Q_{X_{1}X_{2}\ldots X_{K}}^{*}$ with $Q_{X_{1}X_{2}\ldots X_{K-1}}^{*}\otimes Q_{X_{K}|X_{K-1}}^{*}$.
Next, using a similar argument, the first KL divergence term $D_{\text{KL}}(Q_{X_{1}X_{2}\cdots X_{K-1}}\mid\mid P_{X}^{\otimes(K-1)})$
in (\ref{eq: a lower bound on the KL divergence if Markov is not satisfied})
can be similarly lower bounded, showing that $Q_{X_{1}X_{2}\ldots X_{K-1}}^{*}$can
be replaced by $Q_{X_{1}X_{2}\ldots X_{K-2}}^{*}\otimes Q_{X_{K-1}|X_{K-2}}^{*}$
to obtain a lower objective. Continuing repeating this argument in
a recursive fashion results the Markov chain relation any optimal
solution must satisfy. 

Consequently, by the chain rule for the KL divergence
\begin{align}
 & \frac{1}{K}D_{\text{KL}}\left(Q_{X_{1}X_{2}\ldots X_{K}}^{*}\mid\mid P_{X}^{\otimes K}\right)\nonumber \\
 & =\frac{1}{K}\left[D_{\text{KL}}\left(Q_{X_{1}}^{*}\mid\mid P_{X}\right)+\sum_{j=2}^{K}D_{\text{KL}}\left(Q_{X_{j}|X_{j-1}}^{*}\mid\mid P_{X}\mid Q_{X_{j-1}}^{*}\right)\right]\\
 & =\frac{1}{K}\left[D_{\text{KL}}\left(Q_{\tilde{X}_{1}}^{*}\mid\mid P_{X}\right)+\sum_{j=2}^{K}D_{\text{KL}}\left(Q_{\tilde{X}_{2}|\tilde{X}_{1}}^{*}\mid\mid P_{X}\mid Q_{\tilde{X}_{1}}^{*}\right)\right]\\
 & =\left[\frac{1}{K}D_{\text{KL}}\left(Q_{\tilde{X}_{1}}^{*}\mid\mid P_{X}\right)+\frac{K-1}{K}D_{\text{KL}}\left(Q_{\tilde{X}_{2}|\tilde{X}_{1}}^{*}\mid\mid P_{X}\mid Q_{\tilde{X}_{1}}\right)\right].
\end{align}
Moreover, observing (\ref{eq: rate function for expected Bhat bound optimization relaxed}),
we may add the constraint $Q_{\tilde{X}_{1}}=Q_{\tilde{X}_{2}}$ to
the outer minimization, since otherwise the inner constraint $Q_{X_{1}X_{2}}=Q_{X_{2}X_{3}}$,
e.g., would make the problem infeasible. Hence, from all the above,
\begin{equation}
\varphi_{K}(P_{XY})=\min_{Q_{\tilde{X}_{1}\tilde{X}_{2}}:Q_{\tilde{X}_{1}}=Q_{\tilde{X}_{2}}}\left\{ \frac{1}{K}D_{\text{KL}}\left(Q_{\tilde{X}_{1}}\mid\mid P_{X}\right)+\frac{K-1}{K}D_{\text{KL}}\left(Q_{\tilde{X}_{2}|\tilde{X}_{1}}\mid\mid P_{X}\mid Q_{\tilde{X}_{1}}\right)+d(Q_{\tilde{X}_{1}\tilde{X}_{2}})\right\} .\label{eq: rate function for expected Bhat bound optimization relaxed after Markov}
\end{equation}
 Now, by convexity of the KL divergence, it holds that 
\begin{equation}
D_{\text{KL}}\left(Q_{\tilde{X}_{2}|\tilde{X}_{1}}\mid\mid P_{X}\mid Q_{\tilde{X}_{1}}\right)\geq D_{\text{KL}}\left(Q_{\tilde{X}_{1}}\mid\mid P_{X}\right),\label{eq: convexity of the KL divergence for pairs}
\end{equation}
that is, the first KL divergence in (\ref{eq: rate function for expected Bhat bound optimization relaxed after Markov})
is smaller than the second one. Thus, the worst bound is obtained
for $K=4$, that is $\varphi_{K}(P_{XY})\geq\varphi_{4}(P_{XY})$
for all $K\geq4$. Finally, 
\begin{align}
 & \varphi_{4}(P_{XY})\nonumber \\
 & =\min_{Q_{\tilde{X}_{1}\tilde{X}_{2}}:Q_{\tilde{X}_{1}}=Q_{\tilde{X}_{2}}}\left\{ \frac{1}{4}D_{\text{KL}}\left(Q_{\tilde{X}_{1}}\mid\mid P_{X}\right)+\frac{3}{4}D_{\text{KL}}\left(Q_{\tilde{X}_{2}|\tilde{X}_{1}}\mid\mid P_{X}\mid Q_{\tilde{X}_{1}}\right)+d(Q_{\tilde{X}_{1}\tilde{X}_{2}})\right\} \\
 & \trre[\geq,a]\min_{Q_{\tilde{X}_{1}\tilde{X}_{2}}:Q_{\tilde{X}_{1}}=Q_{\tilde{X}_{2}}}\left\{ \frac{1}{4}D_{\text{KL}}\left(Q_{\tilde{X}_{1}}\mid\mid P_{X}\right)+\frac{1}{4}D_{\text{KL}}\left(Q_{\tilde{X}_{2}}\mid\mid P_{X}\right)+\frac{1}{2}D_{\text{KL}}\left(Q_{\tilde{X}_{2}|\tilde{X}_{1}}\mid\mid P_{X}\mid Q_{\tilde{X}_{1}}\right)+d(Q_{\tilde{X}_{1}\tilde{X}_{2}})\right\} \\
 & \trre[=,b]\min_{Q_{\tilde{X}_{1}\tilde{X}_{2}}:Q_{\tilde{X}_{1}}=Q_{\tilde{X}_{2}}}\left\{ \frac{1}{2}D_{\text{KL}}\left(Q_{\tilde{X}_{1}}\mid\mid P_{X}\right)+\frac{1}{2}D_{\text{KL}}\left(Q_{\tilde{X}_{2}|\tilde{X}_{1}}\mid\mid P_{X}\mid Q_{\tilde{X}_{1}}\right)+d(Q_{\tilde{X}_{1}\tilde{X}_{2}})\right\} \\
 & \trre[=,c]\min_{Q_{\tilde{X}_{1}\tilde{X}_{2}}:Q_{\tilde{X}_{1}}=Q_{\tilde{X}_{2}}}\left\{ \frac{1}{2}D_{\text{KL}}\left(Q_{\tilde{X}_{1}\tilde{X}_{2}}\mid\mid P_{X}^{\otimes2}\right)+d(Q_{\tilde{X}_{1}\tilde{X}_{2}})\right\} \\
 & =\psi_{2}(P_{XY}),
\end{align}
where $(a)$ follows using the convexity of the KL divergence, as
in (\ref{eq: convexity of the KL divergence for pairs}) (used with
a factor of $1/4$), $(b)$ follows from the constraint $Q_{\tilde{X}_{1}}=Q_{\tilde{X}_{2}}$,
and $(c)$ follows from the chain rule for KL divergence. Thus, $\varphi_{K}(P_{XY})\geq\psi_{2}(P_{XY})$
for all $K\geq4$. This, combined with the bound $\varphi_{3}(P_{XY})\geq\psi_{2}(P_{XY})$
previously derived completes the proof. 
\end{IEEEproof}
The proof of Theorem \ref{thm: Critical beta for no repeat read regime}
is then as follows:
\begin{IEEEproof}[Proof of Theorem \ref{thm: Critical beta for no repeat read regime}]
Let $F(\pi)$ be the number of fixed points of the permutation $\pi\in S_{M}$,
that is, $F(\pi):=|\{i\in[M]\colon F(i)=i\}|$. So, if $F(\pi)\geq M(1-\xi)$
then $\pi(X^{N})$ is a successful reconstruction of $X^{N}$ with
probability $1$. Hence,
\begin{align}
\mathsf{FP}(\xi) & \trre[\leq,a]\sum_{\pi\in S_{M}:F(\pi)\leq M(1-\xi)}p_{e}[X^{N}\to\pi(X^{N})]\\
 & \trre[\leq,b]\sum_{K=\xi M}^{M}\sum_{\pi\in S_{M}:F(\pi)=M-K}\E\left[e^{-d(X^{N},\pi[X^{N}])}\right]\\
 & \trre[\leq,c]\sum_{K=\xi M}^{M}e^{\frac{K}{\beta}L}\cdot\max_{\pi\in S_{M}:F(\pi)=M-K}\E\left[e^{-d(X^{N},\pi[X^{N}])}\right],
\end{align}
where $(a)$ follows from the union bound, $(b)$ follows from Bhattacharyya's
bound \cite[Sec. 2.3]{viterbi2009principles}, and $(c)$ follows
since the set of permutations which have exactly $M-K$ fixed point
has cardinality of $\binom{M}{K}K!\leq\prod_{j=0}^{K-1}(M-j)\leq M^{K}=e^{\frac{K}{\beta}L}$.
Recall that the Bhattacharyya distance is additive, that is, $d(X^{N},\pi[X^{N}])=\sum_{i\in[M]}d(\boldsymbol{X}(i),\boldsymbol{X}(\pi(i))).$
Now, consider a permutation with $F(\pi)=M-2$ fixed points. This
is a transposition, and since $d(\boldsymbol{X}(i),\boldsymbol{X}(\pi(i))=0$
if $\pi(i)=i$, it follows from Lemma \ref{lem: Bhat bound rate}
that
\[
\E\left[e^{-d(X^{N},\pi[X^{N}])}\right]\leq e^{-2L\cdot\psi_{2}(P_{XY})}.
\]
Similarly, a permutation with $F(\pi)=M-3$ fixed points can only
be a cycle of length $3$, and so it follows again from Lemma \ref{lem: Bhat bound rate}
that
\[
\E\left[e^{-d(X^{N},\pi[X^{N}])}\right]\leq e^{-3L\cdot\psi_{3}(P_{XY})}\leq e^{-3L\cdot\psi_{2}(P_{XY})}.
\]
Next, a permutation with $F(\pi)\leq M-4$ may be either a cycle or
comprised of independent cycles. Suppose that the permutation has
$C$ cycles of lengths $\{K_{j}\}_{j\in[C]}$. By permuting the fragments
of $X^{N}$ if necessary, we may assume WLOG that they are consecutive,
that is, the first cycle includes indices ${\cal I}_{1}:=\{1,2,\ldots,K_{1}\}$,
the second includes ${\cal I}_{2}:=\{K_{1}+1,\ldots,K_{1}+K_{2}\}$
and so on. Since there are $F(\pi)$ fixed points it holds that $\sum_{j\in[C]}K_{j}=M-F(\pi)$.
Thus we may write 
\[
d(X^{N},\pi[X^{N}])=\sum_{j=1}^{C}\sum_{i\in{\cal I}_{j}}d\left(\boldsymbol{X}(i),\boldsymbol{X}(\pi(i))\right),
\]
in which the outer summation is over independent RVs. Then, 
\begin{align}
\E\left[e^{-d(X^{N},\pi[X^{N}])}\right] & \trre[=,a]\prod_{j=1}^{C}\E\left[e^{-\sum_{i\in[M]}d\left(\boldsymbol{X}(i),\boldsymbol{X}(\pi(i))\right)}\right]\\
 & \trre[\leq,b]\prod_{j=1}^{C}e^{-LK_{j}\psi_{2}(P_{XY})}\\
 & \trre[=,c]e^{-(M-F(\pi))L\cdot\psi_{2}(P_{XY})},
\end{align}
where $(a)$ holds by independence, $(b)$ follows from Lemma \ref{lem: Bhat bound rate},
and $(c)$ follows since $\pi$ has $F(\pi)$ fixed points. 

Combining all the above, 
\[
\mathsf{FP}(\xi)\leq\sum_{K=\xi M}^{M}\exp\left[-KL\cdot\left(\psi_{2}(P_{XY})-\frac{1}{\beta}\right)\right].
\]
Now, suppose that $\xi=0$. Then if $K=0$ then the permutation must
be the identity permutation, and this is a perfect reconstruction.
$K=1$ is impossible, since a permutation in $S_{M}$ cannot have
$M-1$ fixed points. Hence, 
\begin{align}
\mathsf{FP}(\xi=0) & \leq\sum_{K=2}^{\infty}\exp\left[-KL\cdot\left(\psi_{2}(P_{XY})-\frac{1}{\beta}\right)\right]\\
 & \trre[=,a]\frac{\exp\left[-2L\cdot\left(\psi_{2}(P_{XY})-\frac{1}{\beta}\right)\right]}{1-\exp\left[-L\cdot\left(\psi_{2}(P_{XY})-\frac{1}{\beta}\right)\right]}\\
 & \trre[\leq,b]2\cdot M^{2(1-\beta\psi_{2}(P_{XY}))},
\end{align}
where $(a)$ is a geometric series, and $(b)$ holds when $1-\exp[-L\cdot(\psi_{2}(P_{XY})-\frac{1}{\beta})]\leq\frac{1}{2}$,
which holds for all $M\geq M_{0}(P_{XY})$. Otherwise, if $\xi>0$
then we may upper bound the sum by $M$ times its maximal term, and
so 
\[
\mathsf{FP}(\xi)\leq\exp\left[-M\log M\cdot\xi\left(\beta\psi_{2}(P_{XY})-1-O\left(\frac{1}{M}\right)\right)\right].
\]
 This completes the proof of the theorem. 
\end{IEEEproof}
We now turn to prove the lower bound in Theorem \ref{thm: tightness of no-repeat}. 
\begin{IEEEproof}[Proof of Theorem \ref{thm: tightness of no-repeat}]
To lower bound the error probability, we first focus on a single
pair of fragments $(i_{1},i_{2})$ and lower bound the probability
that a reconstruction which puts $\boldsymbol{X}(i_{1})$ in the $i_{2}$th
location and $\boldsymbol{X}(i_{2})$ in the $i_{1}$th location is
more likely than the opposite ordering. Concretely, for a pair $(i_{1},i_{2})\in[M]^{2}$
such that $i_{1}<i_{2}$, we let 
\begin{equation}
\tilde{{\cal E}}_{i_{1}i_{2}}:=\left\{ \frac{P_{Y|X}^{\otimes2L}\left[\left(\boldsymbol{Y}(i_{2}),\boldsymbol{Y}(i_{1})\right)\mid\left(\boldsymbol{X}(i_{1}),\boldsymbol{X}(i_{2})\right)\right]}{P_{Y|X}^{\otimes2L}\left[\left(\boldsymbol{Y}(i_{1}),\boldsymbol{Y}(i_{1})\right)\mid\left(\boldsymbol{X}(i_{1}),\boldsymbol{X}(i_{2})\right)\right]}\geq1\right\} ,\label{eq: transposition pairwise error event}
\end{equation}
and the event of interest is defined as the intersection of this event
with the event that the two fragments are different, to wit,
\begin{equation}
{\cal E}_{i_{1}i_{2}}:=\tilde{{\cal E}}_{i_{1}i_{2}}\cap\left\{ \boldsymbol{X}(i_{1})\neq\boldsymbol{X}(i_{2})\right\} .\label{eq: transposition pairwise error event different}
\end{equation}
It should be noted, however, that the event ${\cal E}_{i_{1}i_{2}}$
does not necessarily imply that the ML reconstruction transposes the
pair $(i_{1},i_{2})$, since, for example, placing $\boldsymbol{X}(i_{1})$
in an index different from $i_{2}$ could result larger likelihood.
For notational simplicity, we next assume that $i_{1}=1$ and $i_{2}=2$.
Then, the probability of ${\cal E}_{12}$ equals to the error probability
of the hypothesis testing problem between $H_{0}:(\boldsymbol{X}(1),\boldsymbol{X}(2))=(\tilde{\boldsymbol{X}}(1),\tilde{\boldsymbol{X}}(2))$
and $H_{1}:(\boldsymbol{X}(1),\boldsymbol{X}(2))=(\tilde{\boldsymbol{X}}(2),\tilde{\boldsymbol{X}}(1))$
based on the observations $\boldsymbol{Y}\sim P_{Y|X}^{\otimes2L}(\cdot\mid(\boldsymbol{X}(1),\boldsymbol{X}(2)))$,
when $\tilde{\boldsymbol{X}}(1),\tilde{\boldsymbol{X}}(2)\sim P_{X}^{\otimes L}$,
independently, except whenever $\tilde{\boldsymbol{X}}(1)=\tilde{\boldsymbol{X}}(2)$,
because then the hypothesis testing problem has large error probability,
whereas the reconstruction failure probability is zero. 

Consider first a fixed $\tilde{\boldsymbol{X}}(1)=\tilde{\boldsymbol{x}}(1),\tilde{\boldsymbol{X}}(2)=\tilde{\boldsymbol{x}}(2)$,
and let the probability of erroneously deciding $H_{1}$ when $H_{0}$
is true (resp. deciding $H_{0}$ when $H_{1}$ is true) be $p_{0\to1}(\tilde{\boldsymbol{x}}(1),\tilde{\boldsymbol{x}}(2))$
(resp. $p_{1\to0}(\tilde{\boldsymbol{x}}(1),\tilde{\boldsymbol{x}}(2))$).
By the celebrated result of Shannon, Gallager and Berlekamp \cite[Corollary to Thm. 5]{shannon1967lower1}
the error probability is lower bounded using the Chernoff distance
(\ref{eq: Chernoff distance def}). Specifically, the version in \cite[Problem 10.20(b)]{csiszar2011information}
states that for any $\delta>0$ 
\begin{align}
 & \max\left\{ p_{0\to1}(\tilde{\boldsymbol{x}}(1),\tilde{\boldsymbol{x}}(2)),p_{1\to0}(\tilde{\boldsymbol{x}}(1),\tilde{\boldsymbol{x}}(2))\right\} \nonumber \\
 & \geq\exp\left[-L\cdot\left(\max_{s\in[0,1]}d_{s}\left((\tilde{\boldsymbol{x}}(1),\tilde{\boldsymbol{x}}(2)),(\tilde{\boldsymbol{x}}(2),\tilde{\boldsymbol{x}}(1))\right)+\delta\right)\right]\\
 & =\exp\left[-L\cdot\left(\max_{s\in[0,1]}d_{s}\left(\tilde{\boldsymbol{x}}(1),\tilde{\boldsymbol{x}}(2)\right)+d_{1-s}\left(\tilde{\boldsymbol{x}}(1),\tilde{\boldsymbol{x}}(2)\right)+\delta\right)\right]
\end{align}
holds for all $L\geq L_{0}(\delta,P_{Y|X})$. Now, $s\to d_{s}(\overline{x},\tilde{x})$
is a concave function of $s\in[0,1]$ (its second derivative $\frac{\d^{2}}{\d^{2}s}d_{s}(\overline{x},\tilde{x})$
is the variance of the tilted distribution $\frac{P_{Y|X}^{s}[y\mid\overline{x}]\cdot P_{Y|X}^{1-s}[y\mid\tilde{x}]}{\exp[-d_{s}(\overline{x},\tilde{x})]}$
and hence nonnegative; see \cite[Thm. 5]{shannon1967lower1} and \cite[Proof of Thm. 3.5.1.]{viterbi2009principles}).
Then, $s\to d_{s}\left(\tilde{\boldsymbol{x}}(1),\tilde{\boldsymbol{x}}(2)\right)$
is an average of concave functions and thus concave. Thus, for any
$s\in[0,1]$
\[
\frac{1}{2}d_{s}\left(\tilde{\boldsymbol{x}}(1),\tilde{\boldsymbol{x}}(2)\right)+\frac{1}{2}d_{1-s}\left(\tilde{\boldsymbol{x}}(1),\tilde{\boldsymbol{x}}(2)\right)\leq d_{s/2+(1-s)/2}\left(\tilde{\boldsymbol{x}}(1),\tilde{\boldsymbol{x}}(2)\right)=d\left(\tilde{\boldsymbol{x}}(1),\tilde{\boldsymbol{x}}(2)\right),
\]
that is, the Bhattacharyya distance. We thus get from the above
\[
\max\left\{ p_{0\to1}(\tilde{\boldsymbol{x}}(1),\tilde{\boldsymbol{x}}(2)),p_{1\to0}(\tilde{\boldsymbol{x}}(1),\tilde{\boldsymbol{x}}(2))\right\} \geq\exp\left[-L\cdot\left(2d\left(\tilde{\boldsymbol{x}}(1),\tilde{\boldsymbol{x}}(2)\right)+\delta\right)\right].
\]
We next average this bound over the randomness of $\tilde{\boldsymbol{X}}(1),\tilde{\boldsymbol{X}}(2)$,
while accounting for the requirement that $\tilde{\boldsymbol{X}}(1)\neq\tilde{\boldsymbol{X}}(2)$.
To this end, let $a(\tilde{\boldsymbol{X}}(1),\tilde{\boldsymbol{X}}(2))$
be any integrable function of $\tilde{\boldsymbol{X}}(1),\tilde{\boldsymbol{X}}(2)$
(w.r.t. the probability measure $P_{X}^{\otimes2L}$), that is upper
bounded as $a(\tilde{\boldsymbol{x}}(1),\tilde{\boldsymbol{x}}(2))\le1$
for all $\tilde{\boldsymbol{x}}(1),\tilde{\boldsymbol{x}}(2)\in{\cal X}^{L}$.
Then, 
\begin{align}
 & \E\left[a(\tilde{\boldsymbol{X}}(1),\tilde{\boldsymbol{X}}(2))\cdot\I\{\tilde{\boldsymbol{X}}(1)\neq\tilde{\boldsymbol{X}}(2)\}\right]\nonumber \\
 & =\E\left[a(\tilde{\boldsymbol{X}}(1),\tilde{\boldsymbol{X}}(2))\right]-\E\left[a(\tilde{\boldsymbol{X}}(1),\tilde{\boldsymbol{X}}(2))\cdot\I\{\tilde{\boldsymbol{X}}(1)=\tilde{\boldsymbol{X}}(2)\}\right]\\
 & \geq\E\left[a(\tilde{\boldsymbol{X}}(1),\tilde{\boldsymbol{X}}(2))\right]-\P\left[\tilde{\boldsymbol{X}}(1)=\tilde{\boldsymbol{X}}(2)\right]\\
 & =\E\left[a(\tilde{\boldsymbol{X}}(1),\tilde{\boldsymbol{X}}(2))\right]-e^{-LH_{2}(P_{X})},
\end{align}
where $H_{2}(P_{X})$ is the second-order R\'{e}nyi entropy (the
collision entropy). Hence, using the method of types \cite[Sec. 2.1]{csiszar2011information}
\begin{align}
 & \E\left[\max\left\{ p_{0\to1}(\tilde{\boldsymbol{X}}(1),\tilde{\boldsymbol{X}}(2)),p_{1\to0}(\tilde{\boldsymbol{X}}(1),\tilde{\boldsymbol{X}}(2))\right\} \cdot\I\{\tilde{\boldsymbol{X}}(1)\neq\tilde{\boldsymbol{X}}(2)\}\right]\nonumber \\
 & \geq\E\left[\exp\left[-L\cdot\left(2d\left(\tilde{\boldsymbol{X}}(1),\tilde{\boldsymbol{X}}(2)\right)+\delta\right)\right]\right]-e^{-LH_{2}(P_{X})}\\
 & =\sum_{Q_{\tilde{X}_{1}\tilde{X}_{2}}\in{\cal P}_{L}({\cal X}^{2})}\P\left[(\tilde{\boldsymbol{X}}(1),\tilde{\boldsymbol{X}}(2))\in T_{L}(Q_{\tilde{X}_{1}\tilde{X}_{2}})\right]\cdot\exp\left[-L\cdot\left(2d(Q_{\tilde{X}_{1}\tilde{X}_{2}})+\delta\right)\right]-e^{-LH_{2}(P_{X})}\\
 & \geq\frac{1}{(L+1)^{|{\cal X}|^{2}}}\max_{Q_{\tilde{X}_{1}\tilde{X}_{2}}\in{\cal P}_{L}({\cal X}^{2})}\exp\left[-2L\cdot\left(\frac{1}{2}D_{\text{KL}}(Q_{\tilde{X}_{1}\tilde{X}_{2}}\mid\mid P_{X}^{\otimes2})+d(Q_{\tilde{X}_{1}\tilde{X}_{2}})+\frac{\delta}{2}\right)\right]-e^{-LH_{2}(P_{X})}\\
 & =\exp\left[-2L\cdot\min_{Q_{\tilde{X}_{1}\tilde{X}_{2}}\in{\cal P}_{L}({\cal X}^{2})}\left(\frac{1}{2}D_{\text{KL}}(Q_{\tilde{X}_{1}\tilde{X}_{2}}\mid\mid P_{X}^{\otimes2})+d(Q_{\tilde{X}_{1}\tilde{X}_{2}})+\frac{\delta}{2}-O\left(\frac{|{\cal X}|^{2}\log L}{L}\right)\right)\right]\nonumber \\
 & \hphantom{======}-e^{-LH_{2}(P_{X})}.\label{eq: rate function for lower bound on ML of transposition}
\end{align}
Let 
\[
f(Q_{\tilde{X}_{1}\tilde{X}_{2}}):=\frac{1}{2}D_{\text{KL}}(Q_{\tilde{X}_{1}\tilde{X}_{2}}\mid\mid P_{X}^{\otimes2})+d(Q_{\tilde{X}_{1}\tilde{X}_{2}})
\]
be the objective function involved in the optimization of the above
rate function. Our next goal is to consider the optimization over
this function, which in (\ref{eq: rate function for lower bound on ML of transposition})
is over ${\cal P}_{L}({\cal X}^{2})$. In order to replace this optimization
set with a simpler optimization over the entire probability simplex
${\cal P}({\cal X}^{2})$, it suffices to prove that the function
$Q_{\tilde{X}_{1}\tilde{X}_{2}}\to f(Q_{\tilde{X}_{1}\tilde{X}_{2}})$
is equicontinuous w.r.t. to $Q_{\tilde{X}_{1}\tilde{X}_{2}}$ over
the probability simplex ${\cal P}({\cal X}^{2})$. First, we decompose
the KL divergence as 
\[
D_{\text{KL}}(Q_{\tilde{X}_{1}\tilde{X}_{2}}\mid\mid P_{X}^{\otimes2})=-H(Q_{\tilde{X}_{1}\tilde{X}_{2}})-\E_{Q}\left[\log P_{X}^{\otimes2}(\tilde{X}_{1},\tilde{X}_{2})\right].
\]
Now, first, for any $Q_{\tilde{X}_{1}\tilde{X}_{2}},Q_{\overline{X}_{1}\overline{X}_{2}}\in{\cal P}({\cal X}^{2})$
it holds that \cite[Lemma 2.7]{csiszar2011information}
\[
\left|H(Q_{\tilde{X}_{1}\tilde{X}_{2}})-H(Q_{\overline{X}_{1}\overline{X}_{2}})\right|\leq d_{\text{TV}}\left(Q_{\tilde{X}_{1}\tilde{X}_{2}},Q_{\overline{X}_{1}\overline{X}_{2}}\right)\cdot\log\frac{|{\cal X}|^{2}}{d_{\text{TV}}\left(Q_{\tilde{X}_{1}\tilde{X}_{2}},Q_{\overline{X}_{1}\overline{X}_{2}}\right)},
\]
and, furthermore, 
\begin{align}
 & \left|\E_{Q}\left[\log P_{X}^{\otimes2}(\tilde{X}_{1},\tilde{X}_{2})\right]-\E_{Q}\left[\log P_{X}^{\otimes2}(\overline{X}_{1},\overline{X}_{2})\right]\right|\nonumber \\
 & =\left|\sum_{x_{1},x_{2}\in{\cal X}}\left[Q_{\tilde{X}_{1}\tilde{X}_{2}}(x_{1},x_{2})-Q_{\overline{X}_{1}\overline{X}_{2}}(x_{1},x_{2})\right]\cdot\log\left(P_{X}^{\otimes2}(x_{1},x_{2})\right)\right|\\
 & \leq d_{\text{TV}}\left(Q_{\tilde{X}_{1}\tilde{X}_{2}},Q_{\overline{X}_{1}\overline{X}_{2}}\right)\cdot2\max_{x\in{\cal X}}\left[-\log P_{X}(x)\right].
\end{align}
Second, for the Bhattacharyya distance 
\begin{align}
 & \left|d(Q_{\tilde{X}_{1}\tilde{X}_{2}})-d(Q_{\overline{X}_{1}\overline{X}_{2}})\right|\nonumber \\
 & \leq\left|\sum_{x_{1},x_{2}\in{\cal X}}\left[Q_{\tilde{X}_{1}\tilde{X}_{2}}(x_{1},x_{2})-Q_{\overline{X}_{1}\overline{X}_{2}}(x_{1},x_{2})\right]\cdot d(x_{1},x_{2})\right|\\
 & \leq d_{\text{TV}}\left(Q_{\tilde{X}_{1}\tilde{X}_{2}},Q_{\overline{X}_{1}\overline{X}_{2}}\right)\cdot\max_{x_{1},x_{2}\in{\cal X}}d(x_{1},x_{2}).
\end{align}
Thus, the triangle inequality implies 
\[
\left|f(Q_{\tilde{X}_{1}\tilde{X}_{2}})-f(Q_{\overline{X}_{1}\overline{X}_{2}})\right|\leq d_{\text{TV}}\left(Q_{\tilde{X}_{1}\tilde{X}_{2}},Q_{\overline{X}_{1}\overline{X}_{2}}\right)\cdot\left[\log\frac{|{\cal X}|^{2}}{d_{\text{TV}}\left(Q_{\tilde{X}_{1}\tilde{X}_{2}},Q_{\overline{X}_{1}\overline{X}_{2}}\right)}+c(P_{XY})\right],
\]
where 
\[
c(P_{XY})=2\max_{x\in{\cal X}}\left[-\log P_{X}(x)\right]+\max_{x_{1},x_{2}\in{\cal X}}d(x_{1},x_{2}).
\]
Now, for any given PMF $Q_{\overline{X}_{1}\overline{X}_{2}}\in{\cal P}({\cal X}^{2})$
there exists a type $Q_{\tilde{X}_{1}\tilde{X}_{2}}\in{\cal P}_{L}({\cal X}^{2})$
such that 
\[
d_{\text{TV}}\left(Q_{\tilde{X}_{1}\tilde{X}_{2}},Q_{\overline{X}_{1}\overline{X}_{2}}\right)=\sum_{x_{1},x_{2}\in{\cal X}^{2}}\left|Q_{\tilde{X}_{1}\tilde{X}_{2}}(x_{1},x_{2})-Q_{\overline{X}_{1}\overline{X}_{2}}(x_{1},x_{2})\right|\leq\frac{|{\cal X}|^{2}}{L}.
\]
It further holds that $t\to t\log(1/t)$ is increasing for $t\in[0,e^{-1}]$.
Hence, if $L$ is large enough so that $\frac{|{\cal X}|^{2}}{L}\leq e^{-1}$
and $c(P_{XY})\leq\log L$, it holds that 
\[
\left|f(Q_{\tilde{X}_{1}\tilde{X}_{2}})-f(Q_{\overline{X}_{1}\overline{X}_{2}})\right|\leq\frac{|{\cal X}|^{2}}{L}\left[\log(L)+c(P_{XY})\right]\leq\frac{2|{\cal X}|^{2}\log L}{L}.
\]
So there exists $L_{1}(P_{XY})$ such that for all $L\geq L_{1}(P_{XY})$
it holds that 
\[
\left|f(Q_{\tilde{X}_{1}\tilde{X}_{2}})-f(Q_{\overline{X}_{1}\overline{X}_{2}})\right|\leq\frac{2|{\cal X}|^{2}\log L}{L}.
\]
We may then replace the first exponent in (\ref{eq: rate function for lower bound on ML of transposition})
with 
\begin{equation}
\min_{Q_{\tilde{X}_{1}\tilde{X}_{2}}\in{\cal P}({\cal X}^{2})}\left(\frac{1}{2}D_{\text{KL}}(Q_{\tilde{X}_{1}\tilde{X}_{2}}\mid\mid P_{X}^{\otimes2})+d(Q_{\tilde{X}_{1}\tilde{X}_{2}})+\frac{\delta}{2}-O\left(\frac{|{\cal X}|^{2}\log L}{L}\right)\right),\label{eq: rate function for lower bound on ML of transposition PMFs}
\end{equation}
where now the outer minimization is over $Q_{\tilde{X}_{1}\tilde{X}_{2}}$
that is not necessarily restricted to be a type in ${\cal P}_{L}({\cal X}^{2})$,
but rather any joint PMF in the probability simplex ${\cal P}({\cal X}^{2})$.
Ignoring the terms of $\delta/2$ and the asymptotically vanishing
term, (\ref{eq: rate function for lower bound on ML of transposition PMFs})
is exactly $\psi_{2}(P_{XY})$ defined in (\ref{eq: rate function for expected Bhat bound}).
After taking $\delta\downarrow0$, we obtain the lower bound 
\begin{align}
\P[{\cal E}_{12}] & \geq\E\left[\max\left\{ p_{0\to1}(\tilde{\boldsymbol{X}}(1),\tilde{\boldsymbol{X}}(2)),p_{1\to0}(\tilde{\boldsymbol{X}}(1),\tilde{\boldsymbol{X}}(2))\right\} \cdot\I\{\tilde{\boldsymbol{X}}(1)\neq\tilde{\boldsymbol{X}}(2)\}\right]\\
 & \geq e^{-2L\cdot[\psi_{2}(P_{XY})+o(1)]}-e^{-LH_{2}(P_{X})}\\
 & \trre[\geq,a]e^{-2L\cdot[\psi_{2}(P_{XY})+o(1)]}\\
 & =M^{-2\beta\psi_{2}(P_{XY})+o(1)},\label{eq: a lower bound on the transposition error}
\end{align}
where $(a)$ holds under the assumption $2\psi_{2}(P_{XY})<H_{2}(P_{X})$
assuming that $L$ is sufficiently large $L\geq L_{0}(P_{XY})\vee L_{1}(P_{XY})$.
Note that $\P[{\cal E}_{12}]$ is less than the probability of a transposition
error of $\boldsymbol{X}(i_{1})$ and $\boldsymbol{X}(i_{2})$, which
was upper bounded in the proof of Theorem \ref{thm: Critical beta for no repeat read regime}
as 
\[
\P[{\cal E}_{12}]\leq\P[\tilde{{\cal E}}_{12}]\leq\P\left[\hat{\boldsymbol{X}}(i_{1})=\boldsymbol{X}(i_{2}),\;\hat{\boldsymbol{X}}(i_{2})=\boldsymbol{X}(i_{1})\right]\leq e^{-2L\cdot\psi_{2}(P_{XY})}=M^{-2\beta\psi_{2}(P_{XY})}.
\]
So the bound on ${\cal E}_{12}$ is tight in its polynomial decay
rate, and we let its probability be
\[
p:=\P[\tilde{{\cal E}}_{12}]=\P[{\cal E}_{12}]=M^{-2\beta\psi_{2}(P_{XY})+o(1)}.
\]

We next consider separately the case of $\xi=0$ and $\xi>0$, beginning
with the former. To this end, we will lower bound the failure probability
by the probability of the union $\bigcup_{(i_{1},i_{2})\in[M]^{2}\colon i_{1}<i_{2}}{\cal E}_{i_{1}i_{2}}$,
and to lower bound the probability of this union, we will use de Caen's
inequality \cite{de1997lower}. This inequality requires evaluating
the probability of each event, as well as the probability of intersections
of events ${\cal E}_{i_{1}i_{2}}\cap{\cal E}_{i_{3}i_{4}}$. For single
events, it readily holds from the assumption that the fragments $\{\boldsymbol{X}(i)\}_{i\in\mathbb{N}_{+}}$
are drawn IID and from symmetry that $\P[{\cal E}_{i_{1}i_{2}}]=p$
for any pair $(i_{1},i_{2})\in[M]^{2}$ so that $i_{1}<i_{2}$. For
the probability of an intersection of events, there are two possible
cases. If $i_{1},i_{2},i_{3},i_{4}$ are all distinct then the events
${\cal E}_{i_{1}i_{2}}$ and ${\cal E}_{i_{3}i_{4}}$ are independent,
and so trivially,
\[
\P\left[{\cal E}_{i_{1}i_{2}}\cap{\cal E}_{i_{3}i_{4}}\right]=\P\left[{\cal E}_{i_{1}i_{2}}\right]\cdot\P\left[{\cal E}_{i_{3}i_{4}}\right]=p^{2}.
\]
Otherwise, if $i_{1}=i_{3}$ and $i_{2}\neq i_{4}$, then the events
are dependent, and the probability is larger. We next assume for notational
simplicity that $i_{1}=1,i_{2}=2,i_{4}=4$ and upper bound the probability
$\P[{\cal E}_{12}\cap{\cal E}_{14}]$. First, we remove the constraint
that the fragments are different and upper bound 
\begin{align}
\P[{\cal E}_{12}\cap{\cal E}_{14}] & =\P\left[\tilde{{\cal E}}_{12}\cap\left\{ \boldsymbol{X}(1)\neq\boldsymbol{X}(2)\right\} \cap\tilde{{\cal E}}_{14}\cap\left\{ \boldsymbol{X}(1)\neq\boldsymbol{X}(4)\right\} \right]\\
 & \leq\P[\tilde{{\cal E}}_{12}\cap\tilde{{\cal E}}_{14}].
\end{align}
We thus bound the probability on the right-hand side, as in the proof
of the Bhattacharyya bound \cite[Sec. 2.3]{viterbi2009principles}.
To this end, let $(\boldsymbol{X}(1),\boldsymbol{X}(2),\boldsymbol{X}(4))\sim P_{X}^{\otimes3L}$
and let $(\boldsymbol{Y}(1),\boldsymbol{Y}(2),\boldsymbol{Y}(4))\mid(\boldsymbol{X}(1),\boldsymbol{X}(2),\boldsymbol{X}(4))\sim P_{Y|X}^{\otimes3L}$.
Then,
\begin{align}
 & \P\left[\tilde{{\cal E}}_{12}\cap\tilde{{\cal E}}_{14}\right]\nonumber \\
 & =\E\left[\I\left\{ \frac{\P\left[\boldsymbol{Y}(2),\boldsymbol{Y}(1)\mid\boldsymbol{X}(1),\boldsymbol{X}(2)\right]}{\P\left[\boldsymbol{Y}(1),\boldsymbol{Y}(2)\mid\boldsymbol{X}(1),\boldsymbol{X}(2)\right]}\geq1\right\} \times\I\left\{ \frac{\P\left[\boldsymbol{Y}(4),\boldsymbol{Y}(1)\mid\boldsymbol{X}(1),\boldsymbol{X}(4)\right]}{\P\left[\boldsymbol{Y}(1),\boldsymbol{Y}(4)\mid\boldsymbol{X}(1),\boldsymbol{X}(4)\right]}\geq1\right\} \right]\\
 & \trre[\leq,a]\E\Bigg[\sum_{(\boldsymbol{y}(1),\boldsymbol{y}(2),\boldsymbol{y}(4))\in{\cal Y}^{3L}}P_{Y|X}^{\otimes L}\left[\boldsymbol{y}(1)\mid\boldsymbol{X}(1)\right]\cdot P_{Y|X}^{\otimes L}\left[\boldsymbol{y}(2)\mid\boldsymbol{X}(2)\right]\cdot P_{Y|X}^{\otimes L}\left[\boldsymbol{y}(4)\mid\boldsymbol{X}(4)\right]\nonumber \\
 & \hphantom{==}\times\sqrt{\frac{P_{Y|X}^{\otimes L}\left[\boldsymbol{y}(2)\mid\boldsymbol{X}(1)\right]P_{Y|X}^{\otimes L}\left[\boldsymbol{y}(1)\mid\boldsymbol{X}(2)\right]}{P_{Y|X}^{\otimes L}\left[\boldsymbol{y}(1)\mid\boldsymbol{X}(1)\right]P_{Y|X}^{\otimes L}\left[\boldsymbol{y}(2)\mid\boldsymbol{X}(2)\right]}}\times\sqrt{\frac{P_{Y|X}^{\otimes L}\left[\boldsymbol{y}(4)\mid\boldsymbol{X}(1)\right]P_{Y|X}^{\otimes L}\left[\boldsymbol{y}(1)\mid\boldsymbol{X}(4)\right]}{P_{Y|X}^{\otimes L}\left[\boldsymbol{y}(1)\mid\boldsymbol{X}(1)\right]P_{Y|X}^{\otimes L}\left[\boldsymbol{y}(4)\mid\boldsymbol{X}(4)\right]}}\Bigg]\\
 & =\E\Bigg[\sum_{\boldsymbol{y}(1)}\sqrt{P_{Y|X}^{\otimes L}\left[\boldsymbol{y}(1)\mid\boldsymbol{X}(2)\right]P_{Y|X}^{\otimes L}\left[\boldsymbol{y}(1)\mid\boldsymbol{X}(4)\right]}\nonumber \\
 & \hphantom{==}\sum_{\boldsymbol{y}(2)}\sqrt{P_{Y|X}^{\otimes L}\left[\boldsymbol{y}(2)\mid\boldsymbol{X}(1)\right]P_{Y|X}^{\otimes L}\left[\boldsymbol{y}(2)\mid\boldsymbol{X}(2)\right]}\sum_{\boldsymbol{y}(4)}\sqrt{P_{Y|X}^{\otimes L}\left[\boldsymbol{y}(4)\mid\boldsymbol{X}(4)\right]P_{Y|X}^{\otimes L}\left[\boldsymbol{y}(4)\mid\boldsymbol{X}(1)\right]}\Bigg]\\
 & \trre[=,b]\sum_{Q_{X_{1}X_{2}X_{4}}\in{\cal P}_{L}({\cal X}^{3})}\P\left[(\boldsymbol{X}(1),\boldsymbol{X}(2),\boldsymbol{X}(4))\in T_{L}(Q_{X_{1}X_{2}X_{4}})\right]\nonumber \\
 & \hphantom{==}\times\exp\left[-L\cdot\left(d_{P_{Y|X}}(Q_{X_{2}X_{4}})+d_{P_{Y|X}}(Q_{X_{1}X_{2}})+d_{P_{Y|X}}(Q_{X_{4}X_{1}})\right)\right]\\
 & \trre[\leq,c]\left|{\cal P}_{L}({\cal X}^{3})\right|\nonumber \\
 & \hphantom{==}\times\max_{Q_{X_{1}X_{2}X_{4}}\in{\cal P}_{L}({\cal X}^{3})}\exp\left[-L\cdot\left(D_{\text{KL}}(Q_{X_{1}X_{2}X_{4}}\mid\mid P_{X}^{\otimes3})+d_{P_{Y|X}}(Q_{X_{2}X_{4}})+d_{P_{Y|X}}(Q_{X_{1}X_{2}})+d_{P_{Y|X}}(Q_{X_{4}X_{1}})\right)\right]\\
 & \trre[\leq,d]\exp\left[-L\cdot\left(3\psi_{3}(P_{XY})-\frac{|{\cal X}|^{3}\log(L+1)}{L}\right)\right]\\
 & =M^{-3\beta\psi_{3}(P_{XY})+o(1)}\\
 & \trre[\leq,e]M^{-3\beta\psi_{2}(P_{XY})+o(1)}=:q,\label{eq: upper bound on intersection of transposition errors connected}
\end{align}
where $(a)$ follows from the standard Bhattacharyya bound technique
of bounding 
\[
\I\left\{ \frac{\P\left[\boldsymbol{Y}(2),\boldsymbol{Y}(1)\mid\boldsymbol{X}(1),\boldsymbol{X}(2)\right]}{\P\left[\boldsymbol{Y}(1),\boldsymbol{Y}(2)\mid\boldsymbol{X}(1),\boldsymbol{X}(2)\right]}\geq1\right\} \leq\sqrt{\frac{\P\left[\boldsymbol{Y}(2),\boldsymbol{Y}(1)\mid\boldsymbol{X}(1),\boldsymbol{X}(2)\right]}{\P\left[\boldsymbol{Y}(1),\boldsymbol{Y}(2)\mid\boldsymbol{X}(1),\boldsymbol{X}(2)\right]}},
\]
$(b)$ follows since the Bhattacharyya distance between $\boldsymbol{X}(1)$
and $\boldsymbol{X}(2)$ only depends on their joint type $Q_{X_{1}X_{2}}$
(and similarly for the other Bhattacharyya distances), $(c)$ follows
from the method of types (the upper bound of the probability of a
type class \cite[Lemma 2.3]{csiszar2011information}), $(d)$ follows
from the type counting lemma \cite[Lemma 2.2]{csiszar2011information}
and the definition of $\psi_{K}(P_{XY})$ in (\ref{eq: rate function for expected Bhat bound}),
and $(e)$ holds since as was shown in (\ref{eq: psi_3 is larger than psi_2}),
it holds that $\psi_{3}(P_{XY})\geq\psi_{2}(P_{XY})$. 

Using the above bounds, we may lower bound the failure probability
as
\begin{align}
\mathsf{FP}(\xi=0) & \geq\P\left[\bigcup_{(i_{1},i_{2})\in[M]^{2}\colon i_{1}<i_{2}}{\cal E}_{i_{1}i_{2}}\right]\\
 & \trre[\geq,a]\sum_{(i_{1},i_{2})\in[M]^{2}\colon i_{1}<i_{2}}\frac{\P^{2}\left[{\cal E}_{i_{1}i_{2}}\right]}{\sum_{(i_{3},i_{4})\in[M]^{2}\colon i_{3}<i_{4}}\P\left[{\cal E}_{i_{1}i_{2}}\cap{\cal E}_{i_{3}i_{4}}\right]},\label{eq: lower bound xi=00003Do de Caen first step}
\end{align}
where $(a)$ follows from de Caen's inequality \cite{de1997lower}.
The sum's denominator is bounded for any given $(i_{1},i_{2})$ as
follows. For the single term $(i_{3},i_{4})=(i_{1},i_{2})$ 
\[
\P\left[{\cal E}_{i_{1}i_{2}}\cap{\cal E}_{i_{3}i_{4}}\right]=\P\left[{\cal E}_{i_{1}i_{2}}\right]=p.
\]
For the terms in which either $i_{1}=i_{3}$ or $i_{2}=i_{4}$ it
holds that 
\[
\P\left[{\cal E}_{i_{1}i_{2}}\cap{\cal E}_{i_{3}i_{4}}\right]\leq q,
\]
given in (\ref{eq: upper bound on intersection of transposition errors connected}).
There are less than $2M$ such pairs of pairs. Finally, for the terms
in which $i_{1},i_{2},i_{3},i_{4}$ are all distinct it holds that
\[
\P\left[{\cal E}_{i_{1}i_{2}}\cap{\cal E}_{i_{3}i_{4}}\right]=p^{2}.
\]
There are less than $M^{2}$ such terms. Hence, (\ref{eq: lower bound xi=00003Do de Caen first step})
may be further lower bounded as 
\begin{align}
\mathsf{FP}(\xi=0) & \geq\sum_{(i_{1},i_{2})\in[M]^{2}\colon i_{1}<i_{2}}\frac{p^{2}}{p+qM+M^{2}p^{2}}\\
 & \ge\frac{1}{3}\sum_{(i_{1},i_{2})\in[M]^{2}\colon i_{1}<i_{2}}p\wedge\frac{p^{2}}{2qM}\wedge\frac{p^{2}}{M^{2}p^{2}}\\
 & \trre[\geq,a]\frac{1}{24}\cdot\left(M^{2}p\wedge M\frac{p^{2}}{q}\wedge1\right)\\
 & \trre[\geq,b]M^{o(1)}\cdot\left(M^{2(1-\beta\psi_{2}(P_{XY}))}\wedge M^{1-\beta\psi_{2}(P_{XY})}\wedge1\right)\\
 & \trre[=,c]M^{2-2\beta\psi_{2}(P_{XY})+o(1)},
\end{align}
where $(a)$ follows since there are $\frac{M(M-1)}{2}\geq\frac{M^{2}}{4}$
pairs $(i_{1},i_{2})\in[M]^{2}$ such that $i_{1}<i_{2}$ (assuming
trivially that $M\geq2)$, $(b)$ follows by the definition of $p$
in (\ref{eq: a lower bound on the transposition error}) and the definition
of $q=M^{-3\beta\psi_{2}(P_{XY})+o(1)}$ in (\ref{eq: upper bound on intersection of transposition errors connected}),
$(c)$ holds by the assumption of the theorem that $\beta>\frac{1}{\psi_{2}(P_{XY})}$.
This completes the proof of the bound for $\xi=0$.

We now prove the bound for $\xi>0$. Let $I:=\{\{i_{j},i_{j}'\}\}_{j\in[\xi M/2]}$
be a set of $\xi M/2$ unordered pairs of unique indices in $[M]$,
that is $\cap_{j\in[\xi M/2]}\{i_{j}\cap i_{j}'\}=\emptyset$.\footnote{As mentioned we ignore integer constraints, as they are inconsequential
to the final result, and thus assume that $\xi M/2$ is integer.} Consider the event of $\frac{\xi M}{2}$ transposition replacements
of correct likelihood order between pairs of fragments in $I$, that
is 
\[
{\cal E}_{I}:=\bigcap_{j=1}^{\xi M/2}\left\{ {\cal E}_{i_{j},i_{j}'}\right\} 
\]
using the definition of the event ${\cal E}_{i_{1}i_{2}}$ in (\ref{eq: transposition pairwise error event different}).
Then,
\begin{align}
\P[{\cal E}_{I}] & \trre[=,a]\prod_{j=1}^{\xi M/2}\P\left[{\cal E}_{i_{j},i_{j}'}\right]\\
 & \trre[\geq,b]e^{-\xi ML\cdot[\psi_{2}(P_{XY})+o(1)]}\\
 & \trre[=,c]e^{-\xi M\log M\cdot[\beta\psi_{2}(P_{XY})+o(1)]},\label{eq: lower bound on multiple transpositions}
\end{align}
where $(a)$ follows since disjoint pairwise transpositions are independent
events, $(b)$ follows from the lower bound on $\P[{\cal E}_{i_{1}i_{2}}]$
in (\ref{eq: a lower bound on the transposition error}), and $(c)$
from $L=\beta\log M$. 
\end{IEEEproof}

\section{Proofs for the Repeating-Fragments Regime with Nonnegative Distortion
\label{sec: proofs repeating}}

We first prove Prop. \ref{prop: number of possible reconstructions}.
Recall that $G\sim\text{Multinomial}(M;(p_{1},p_{2},\ldots,p_{M^{\beta}}\})$
where $M$ is the fixed number of fragments. Consider the random vector
$\tilde{G}=(\tilde{G}(1),\ldots,\tilde{G}(M^{\beta}))$, which has
the same dimension as $G$, yet each of its components is distributed
$\tilde{G}(j)\sim\text{Poisson}(Mp_{j})$, where $p_{j}:=\P[X^{L}=a_{j}]$
is the probability of $a_{j}\in{\cal X}^{L}$, the $j$th letter in
${\cal X}^{L}$, and where the components of $\tilde{G}$ are \emph{independent}
(unlike those of $G$). By construction, the expected value of each
coordinate in $G$ and $\tilde{G}$ is equal, and given by $\E[\tilde{G}(j)]=\E[G(j)]=Mp_{j}$.
We recall that ``Poissonization of the multinomial'' effect (see
\cite[Sec. 5.4]{mitzenmacher2017probability} for the case $P_{X}$
is uniform and $\{p_{j}\}_{j\in[M^{\beta}]}$ are all equal. This
has a straightforward extension to non-uniform probabilities, see,
e.g., \cite[Lecture 11]{Seshadhri2020lecture}). 
\begin{fact}[Poissonization of the multinomial distribution]
\label{fact: Poissonization of the multinomial distribution}Let
$\tilde{M}\sim\text{\emph{Poisson}}(M)$, and let $\tilde{G}$ be
a random vector such that $\tilde{G}\sim\text{\emph{Multinomial}}(\tilde{M},(p_{1},p_{2},\ldots p_{M^{\beta}}\})$
conditioned on $\tilde{M}$, where $\sum_{j\in[M^{\beta}]}p_{j}=1$
and $p_{j}>0$. Then, $\{\tilde{G}(j)\}_{j\in[M^{\beta}]}$ are statistically
independent and $\tilde{G}(j)\sim\text{\emph{Poisson}}(Mp_{j})$ (unconditioned
on $\tilde{M}$). 
\end{fact}
Fact \ref{fact: Poissonization of the multinomial distribution} can
be verified by spelling out the conditional PMF of $\tilde{G}$ conditioned
on $\tilde{M}$ \cite[Thm. 5.6]{mitzenmacher2017probability} in case
$\{p_{j}\}$ are all equal, and easily extended to the non-uniform
case (e.g., \cite[Lecture 11, Thm. 3.2]{Seshadhri2020lecture}). The
following then follows from \cite[Corollary 5.9]{mitzenmacher2017probability}:
\begin{lem}
\label{lem: Poissonization of events}Let $G\sim\text{\emph{Multinomial}}(M,(p_{1},p_{2},\ldots p_{M^{\beta}}\})$,
and let $\tilde{G}$ be an independent Poisson vector of the same
dimension so that $\E[\tilde{G}(j)]=\E[G(j)]=Mp_{j}$. Then, for any
event ${\cal E}$
\[
\P\left[G\in{\cal E}\right]\leq\sqrt{eM}\cdot\P\left[\tilde{G}\in{\cal E}\right].
\]
\end{lem}
We will also need the following results on Poisson RVs. The first
one is a standard Chernoff bound for Poisson RVs, and the second one
is the aforementioned concentration inequality for Lipschitz functions
of Poisson RVs. 
\begin{lem}[{Chernoff's bound for Poisson RVs \cite[Theorem 5.4]{mitzenmacher2017probability}}]
\label{lem: Chernoff for Poisson rvs}For $W\sim\text{\emph{Poisson}}(\lambda)$
it holds that
\begin{equation}
\P\left[W\geq\alpha\E[W]\right]\leq e^{-\lambda}\left(\frac{e}{\alpha}\right)^{\alpha\lambda}\leq\left(\frac{e}{\alpha}\right)^{\alpha\lambda}=e^{-\alpha\lambda\log(\alpha/e)}\leq e^{-\alpha\lambda},\label{eq: Poisson Chernoff general}
\end{equation}
where the rightmost inequality holds for any $\alpha>3e\approx8.15$. 
\end{lem}
\begin{lem}[Poisson concentration of Lipschitz functions, a variant of \cite{bobkov1998modified,kontoyiannis2006measure}]
\label{lem: Poisson concentration}Let $W\sim\text{\emph{Poisson}}(\lambda)$,
and assume that $f$ is $1$-Lipschitz, that is, $|Df(w)|=|f(w+1)-f(w)|\leq1$
for all $w\in\mathbb{N}_{+}$. Then, for any $t>0$
\begin{equation}
\P\left[f(W)-\E[f(W)]>t\right]\leq\exp\left[-\frac{t^{2}}{16\lambda+3t}\right].\label{eq: Poisson concentration}
\end{equation}
\end{lem}
\begin{IEEEproof}
Under the conditions of the lemma
\begin{align}
 & \P\left[f(W)-\E[f(W)]>t\right]\nonumber \\
 & \trre[\leq,a]\exp\left[-\frac{t}{4}\log\left(1+\frac{t}{2\lambda}\right)\right]\\
 & \trre[=,b]\exp\left[-\frac{\lambda}{2}u\log(1+u)\right]\\
 & \trre[\leq,c]\exp\left[-\frac{\lambda u^{2}}{4(1+u/3)}\right],
\end{align}
where $(a)$ is the Bobkov and Ledoux's bound \cite[Prop. 10]{bobkov1998modified}
\cite{kontoyiannis2006measure}, $(b)$ follows by setting $u:=\frac{t}{2\lambda}$,
$(c)$ follows from
\begin{align}
u\log\left(1+u\right) & =(1+u)\log(1+u)-u+u-\log(1+u)\\
 & \trre[\geq,*]\frac{u^{2}}{2(1+u/3)}+u-\log(1+u)\\
 & \trre[\geq,**]\frac{u^{2}}{2(1+u/3)},
\end{align}
where $(*)$ was stated in \cite[Exercise 2.8]{boucheron2013concentration},
and $(**)$ follows from $u\geq\log(1+u)$ for $u\geq0$. The result
follows by re-substituting $u=\frac{t}{2\lambda}$, and performing
a (minor) numerical relaxation.
\end{IEEEproof}
We may now prove Prop. \ref{prop: number of possible reconstructions}. 
\begin{IEEEproof}[Proof of Prop. \ref{prop: number of possible reconstructions}]
The entropy upper bound on the multinomial coefficient implies that
it surely holds that\footnote{See e.g., \cite[Lemma 17.5.1]{cover2012elements} for the binomial
coefficient; the extension to multinomial is straightforward and well
known.} 
\begin{align}
C(X^{N}):=\frac{1}{M}\log\left|{\cal A}_{L}(X^{N})\right| & =\frac{1}{M}\log\binom{M}{G(1),G(2),\ldots,G(M^{\beta})}\\
 & \trre[\leq,a]H\left(\frac{G(1)}{M},\frac{G(2)}{M},\ldots,\frac{G(M^{\beta})}{M}\right)\\
 & =:-\sum_{j\in[M^{\beta}]}\frac{G(j)}{M}\log\frac{G(j)}{M}.
\end{align}
Let us denote
\[
\tilde{C}:=\sum_{j\in[M^{\beta}]}-\frac{\tilde{G}(j)}{M}\log\frac{\tilde{G}(j)}{M},
\]
where $\tilde{G}(j)\sim\text{Poisson}(M\cdot\P[X^{L}=a_{j}])$, that
is $\E[\tilde{G}(j)]=\E[G(j)]$. Let $\boldsymbol{X}\sim P_{X}^{\otimes L}$
be a random length-$L$ fragment. By Jensen's inequality for the function
$f(t):=-t\log t$ in $\mathbb{R}_{+}$, 
\begin{align}
\E[\tilde{C}] & =\sum_{j\in[M^{\beta}]}\E\left[-\frac{\tilde{G}(j)}{M}\log\frac{\tilde{G}(j)}{M}\right]\\
 & \leq\sum_{j\in[M^{\beta}]}-\frac{\E[\tilde{G}(j)]}{M}\log\frac{\E[\tilde{G}(j)]}{M}\\
 & =-\sum_{j\in[M^{\beta}]}\P[\boldsymbol{X}=a_{j}]\cdot\log\P[\boldsymbol{X}=a_{j}]\\
 & =H(\boldsymbol{X})\\
 & =L\cdot H(P_{X}).
\end{align}
Hence, for any $\eta>0$, the Poissonization effect of Lemma \ref{lem: Poissonization of events}
implies that 
\begin{align}
 & \P\left[C(X^{N})\geq L\cdot H(P_{X})+\eta\log M\right]\nonumber \\
 & \leq\P\left[C(X^{N})\geq\E[\tilde{C}]+\eta\log M\right]\\
 & \leq e\sqrt{M}\cdot\P\left[\tilde{C}\geq\E[\tilde{C}]+\eta\log M\right].\label{eq: poissonization for log cardinality}
\end{align}
We next bound the concentration of $\tilde{C}$ above its expected
value $\E[\tilde{C}]$. We first derive a bound which is effective
in the regime $\beta\in(0,2)$. To this end, we would like to invoke
the concentration inequality of Lipschitz functions of Poisson RVs
by Bobkov and Ledoux \cite{bobkov1998modified,kontoyiannis2006measure},
see Lemma \ref{lem: Poisson concentration}. However, writing $\tilde{C}:=\sum_{j\in[M^{\beta}]}f(\tilde{G}(j))$
for $f(g):=-\frac{g}{M}\log\frac{g}{M}$, it is apparent that this
is not a Lipschitz function on $\mathbb{N}_{+}$ since $f(g)$ has
unbounded derivative for $g\uparrow\infty$. To address this, we first
consider a Lipschitz approximation to $f(g)$ given by $f^{+}(g)=(f(g))_{+}$,
and establish the concentration of $\tilde{C}^{+}=\sum_{j\in[M^{\beta}]}f^{+}(\tilde{G}(j))$
to its expected value $\E[\tilde{C}^{+}]$. Afterwards, we show that
$\E[\tilde{C}^{+}]$ is close to $\E[\tilde{C}]$. It can be easily
verified that the discrete derivative satisfies
\[
\left|f^{+}(g+1)-f^{+}(g)\right|\leq\frac{\log M}{M},
\]
that is, $f^{+}(g)$ is a $(\frac{\log M}{M})$-Lipschitz continuous
function. We first consider the tail behavior of each of the terms
defining $\tilde{C}^{+}$. Let $j\in[M^{\beta}]$ be given. Then,
for any $\eta>0$
\begin{align}
 & \P\left[f^{+}\left(\tilde{G}(j)\right)-\E\left[f^{+}\left(\tilde{G}(j)\right)\right]>\eta\log M\right]\nonumber \\
 & \trre[=,a]\P\left[\hat{f}^{+}\left(\tilde{G}(j)\right)-\E\left[\hat{f}^{+}\left(\tilde{G}(j)\right)\right]>\eta M\right]\\
 & \trre[\leq,b]\exp\left[-\frac{M^{2}\eta^{2}}{16\cdot\E[\tilde{G}(j)]+3M\eta}\right],\label{eq: Concentration via Poisson Lipschitz}
\end{align}
where $(a)$ is obtained by setting $\hat{f}^{+}(g):=\frac{Mf^{+}(g)}{\log M}$
and noting that $\hat{f}^{+}(g)$ is a $1$-Lipschitz continuous function,
$(b)$ is obtained by Poisson concentration of Lipschitz functions,
as stated in Lemma \ref{lem: Poisson concentration}. Since $-\hat{f}^{+}(g)$
is also a $1$-Lipschitz continuous function, an analogous bound holds
for the left tail. Denoting for brevity 
\[
Z(j):=\hat{f}^{+}\left(\tilde{G}(j)\right)-\E\left[\hat{f}^{+}\left(\tilde{G}(j)\right)\right],
\]
we note that if 
\[
M\eta=\sqrt{32\cdot\E[\tilde{G}(j)]t}+6t
\]
then 
\begin{align}
\frac{M^{2}\eta^{2}}{16\cdot\E[\tilde{G}(j)]+3M\eta} & \geq\frac{M^{2}\eta^{2}}{2\left(16\cdot\E[\tilde{G}(j)]\vee3M\eta\right)}\\
 & =\frac{M^{2}\eta^{2}}{32\cdot\E[\tilde{G}(j)]}\wedge\frac{M\eta}{6}\\
 & \geq\frac{32\cdot\E[\tilde{G}(j)]t}{32\cdot\E[\tilde{G}(j)]}\wedge\frac{6t}{6}\\
 & \geq t,
\end{align}
 and so (\ref{eq: Concentration via Poisson Lipschitz}) implies that
\[
\P\left[|Z(j)|\leq\sqrt{32\cdot\E[\tilde{G}(j)]t}+6t\right]\leq2e^{-t}
\]
holds for any $t\geq0$. Hence, the two statements of \cite[Thm. 2.3]{boucheron2013concentration}
together imply that $Z(j)$ is a sub-gamma random variable with variance
proxy $4(128\cdot\E[\tilde{G}(j)]+576)$ and scale parameter $48$.
Since $\E[\tilde{G}(j)]=M\cdot\P[X^{L}=a_{j}]$, there exists a numerical
constant $c_{1}>0$ so that
\begin{align}
\sum_{j\in[M^{\beta}]}\E[Z^{2}(j)] & \leq512\sum_{j\in[M^{\beta}]}\left(\E[\tilde{G}(j)]+2\cdot(24)^{2}\right)\\
 & \leq c_{1}(M+M^{\beta}).
\end{align}
Furthermore, $|Z(j)|\leq2\max_{t\in[0,1]}-t\log t\leq3c_{2}$ for
some numerical constant $c_{2}>0$ (depending on the choice of base
for the logarithm), and so $Z(j)$ satisfies Bernstein's condition
with sum of second moments $\sum_{j\in[M^{\beta}]}\E[Z(j)^{2}]\leq c_{1}(M+M^{\beta})$
and scale constant $c_{2}$. Since $\{Z(j)\}_{j\in[M^{\beta}]}$ are
independent and centered, Bernstein's inequality \cite[Corollary 2.11]{boucheron2013concentration}
then results 
\[
\P\left[\sum_{j\in[M^{\beta}]}Z(j)\geq r\right]\leq\exp\left[-\frac{r^{2}}{2(c_{1}M+c_{1}M^{\beta}+c_{2}r)}\right].
\]
Setting $r=M\eta$, while restricting now that $\eta\in(0,1)$, then
results 
\begin{align}
 & \P\left[\tilde{C}^{+}\geq\E[\tilde{C}^{+}]+\eta\log M\right]\nonumber \\
 & =\P\left[\sum_{j\in[M^{\beta}]}f^{+}\left(\tilde{G}(j)\right)-\E\left[f^{+}\left(\tilde{G}(j)\right)\right]>\eta\log M\right]\\
 & =\P\left[\sum_{j\in[M^{\beta}]}\hat{f}^{+}\left(\tilde{G}(j)\right)-\E\left[\hat{f}^{+}\left(\tilde{G}(j)\right)\right]>M\eta\right]\\
 & \leq\exp\left[-\frac{M^{2}\eta^{2}}{2(c_{1}M+c_{1}M^{\beta}+c_{2}M\eta)}\right]\\
 & \leq\exp\left[-M\cdot\min\left\{ \frac{\eta^{2}}{4c_{1}},\frac{M^{1-\beta}\eta}{c_{1}},\frac{\eta}{c_{2}}\right\} \right]\\
 & \leq\exp\left[-c_{3}\cdot\eta^{2}M^{1\vee(2-\beta)}\right],\label{eq: final concentration of C_Tilde_plus}
\end{align}
for some numerical constant $c_{3}>0$. 

Next, we bound the absolute difference $\E[\tilde{C}^{+}]$ and $\E[\tilde{C}]$.
Note that each of them is comprised of $M^{\beta}$ terms, and as
before, we first focus on a single term $j\in[M^{\beta}]$. For brevity,
we let $\tilde{G}\sim\text{Poisson}(\lambda)$, where $\lambda\leq\frac{M}{10}\leq M-1$
is assumed. Then, 
\begin{align}
 & \E\left[f^{+}(\tilde{G})\right]-\E\left[f(\tilde{G})\right]\nonumber \\
 & =\E\left[\frac{\tilde{G}}{M}\log\frac{\tilde{G}}{M}\cdot\I\{\tilde{G}\geq M\}\right]\\
 & =\sum_{\tilde{g}=M+1}^{\infty}\P[\tilde{G}=\tilde{g}]\frac{\tilde{g}}{M}\log\frac{\tilde{g}}{M}\\
 & =\sum_{\tilde{g}=M+1}^{\infty}\frac{\lambda^{\tilde{g}}e^{-\lambda}}{\tilde{g}!}\frac{\tilde{g}}{M}\log\frac{\tilde{g}}{M}\\
 & =\frac{\lambda}{M}\sum_{\tilde{g}=M+1}^{\infty}\frac{\lambda^{\tilde{g}-1}e^{-\lambda}}{(\tilde{g}-1)!}\log\frac{\tilde{g}}{M}\\
 & \trre[=,b]\frac{\lambda}{M}\sum_{\overline{g}=M}^{\infty}\frac{\lambda^{\overline{g}}e^{-\lambda}}{\overline{g}!}\log\frac{\overline{g}+1}{M}\\
 & =\frac{\lambda}{M}\E\left[\log\frac{\tilde{G}+1}{M}\I\{\tilde{G}\geq M\}\right]\\
 & \trre[\leq,c]\frac{\lambda}{M}\E\left[\frac{(\tilde{G}+1-M)}{M}\cdot\I\{\tilde{G}\geq M\}\right]\\
 & \trre[\leq,d]\frac{\lambda}{M}\E\left[\frac{(\tilde{G}-\lambda)}{M}\cdot\I\{\tilde{G}\geq M\}\right]\\
 & \trre[\leq,e]\frac{\lambda}{M}\sqrt{\E\frac{(\tilde{G}-\lambda)^{2}}{M^{2}}}\cdot\sqrt{\P\left(\tilde{G}\geq M\right)}\\
 & =\frac{\lambda^{3/2}}{M^{2}}\sqrt{\P\left(\tilde{G}\geq M\right)}\\
 & \trre[\leq,f]1\cdot e^{-M/2},\label{eq: difference between expectations in f and f+-1}
\end{align}
where $(a)$ follows since $f^{+}(g)=(f(g))_{+}$, $(b)$ is using
the change of variables $\overline{g}=\tilde{g}-1$, $(c)$ follows
from $\log(t)\leq t-1$ for $t\geq1$, $(d)$ holds since $\lambda\le M-1\leq M$
was assumed, $(e)$ follows from the Cauchy-Schwarz inequality, and
$(f)$ follows from Lemma \ref{lem: Chernoff for Poisson rvs} and
the assumption that $\lambda\leq\frac{M}{10}$. 

Now, under the assumption that $H(P_{X})>0$ it must hold that $\max_{x\in{\cal X}}P_{X}(x)<1$.
Hence, 
\[
\E[\tilde{G}(j)]=M\cdot\P[X^{L}=a_{j}]=M(\max P_{X}(x))^{L}=o(M)
\]
and we may use the approximation of (\ref{eq: difference between expectations in f and f+-1})
assuming that $M\geq M_{0}(P_{X},\beta)$ is sufficiently large. Then,
\begin{equation}
\left|\E[\tilde{C}]-\E[\tilde{C}^{+}]\right|\leq\sum_{j\in[M^{\beta}]}\left|\E\left[f^{+}\left(\tilde{G}(j)\right)\right]-\E\left[f\left(\tilde{G}(j)\right)\right]\right|\leq M^{\beta}e^{-M/2}\leq e^{-M/4}\label{eq: difference betwen expected of Ctilde and Ctilde_plus}
\end{equation}
for all $M\geq M_{1}(P_{X},\beta)$ sufficiently large. Returning
to (\ref{eq: poissonization for log cardinality}) we obtain 
\begin{align}
 & \P\left[C(X^{N})\geq L\cdot H(P_{X})+\eta\log M\right]\nonumber \\
 & \leq e\sqrt{M}\cdot\P\left[\tilde{C}\geq\E[\tilde{C}]+\eta\log M\right]\\
 & \trre[\leq,a]e\sqrt{M}\cdot\P\left[\tilde{C}^{+}\geq\E[\tilde{C}]+\eta\log M\right]\\
 & \trre[\leq,b]e\sqrt{M}\cdot\P\left[\tilde{C}^{+}\geq\E[\tilde{C}^{+}]+\eta\log M-e^{-M/4}\right]\\
 & \trre[\leq,c]e\sqrt{M}\cdot\P\left[\tilde{C}^{+}\geq\E[\tilde{C}^{+}]+\frac{\eta}{2}\log M\right]\\
 & \trre[\leq,d]e\sqrt{M}\cdot\exp\left[-c_{4}\cdot\eta^{2}M^{1\vee(2-\beta)}\right]\\
 & \trre[\leq,e]\exp[-c_{5}\cdot\eta^{2}M^{1\vee(2-\beta)}],
\end{align}
where $(a)$ follows since $\tilde{C}^{+}\geq\tilde{C}$, $(b)$ follows
from (\ref{eq: difference betwen expected of Ctilde and Ctilde_plus}),
$(c)$ holds for all $M\geq M_{2}(\eta)$ sufficiently large, $(d)$
holds from (\ref{eq: final concentration of C_Tilde_plus}), and $(e)$
holds for some numerical constant $c_{5}>0$ and all $M\geq M_{3}(P_{X},\beta,\eta)$
sufficiently large. The result then follows for all $M\geq M_{0}\vee M_{1}\vee M_{2}\vee M_{3}$. 

The bound derived above is non-trivial only for $\beta\in(0,2)$ Next,
we derive a different bound on the one-sided concentration of $\tilde{C}$
above its expected value $\E[\tilde{C}]$ in (\ref{eq: poissonization for log cardinality}),
which is effective for any $\beta>0$. Consider the events 
\[
{\cal F}(j):=\left\{ \tilde{G}(j)\geq M\right\} .
\]
As before, under the assumption that $H(P_{X})>0$ it must hold that
$\E[\tilde{G}(j)]\leq\frac{M}{10}$ for all $M\geq M_{0}(P_{X})$.
Hence, Lemma \ref{lem: Chernoff for Poisson rvs} implies that 
\[
\P\left[{\cal F}(j)\right]=\P\left[\tilde{G}(j)\geq\frac{\E\left[\tilde{G}(j)\right]}{\P[X^{L}=a_{j}]}\right]\leq e^{-M}
\]
(using $\alpha=1/\P[X^{L}=a_{j}]\geq3e$). Letting ${\cal F}=\bigcap_{j\in[M^{\beta}]}{\cal F}(j)$,
the union bound implies that 
\[
\P[{\cal F}]\leq M^{\beta}e^{-M}=e^{-M\left(1-\frac{\beta\log M}{M}\right)},
\]
and the probability in (\ref{eq: poissonization for log cardinality})
is then bounded as 
\begin{align}
 & \P\left[\tilde{C}\geq\E[\tilde{C}]+\eta\log M\right]\nonumber \\
 & \leq\P\left[\left\{ \tilde{C}\geq\E[\tilde{C}]+\eta\log M\right\} \cap{\cal F}^{c}\right]+\P\left[{\cal F}\right]\\
 & =\P\left[\left\{ \tilde{C}\geq\E[\tilde{C}]+\eta\log M\right\} \cap{\cal F}^{c}\right]+e^{-M\left(1-\frac{\beta\log M}{M}\right)}.\label{eq: bound on C tilde concentration with F event}
\end{align}
We further upper bound the first probability. Under ${\cal F}^{c}$,
it holds that $\tilde{G}(j)\leq M$, the argument of $f(t)$ is less
than $1$, and $f(t)\geq0$. Thus, 
\begin{align}
 & \P\left[\left\{ \tilde{C}\geq\E[\tilde{C}]+\eta\log M\right\} \cap{\cal F}^{c}\right]\nonumber \\
 & =\P\left[\left\{ \sum_{j\in[M^{\beta}]}f\left(\tilde{G}(j)\right)\geq\E\left[\sum_{j\in[M^{\beta}]}f\left(\tilde{G}(j)\right)\right]+\eta\log M\right\} \cap{\cal F}^{c}\right]\\
 & =\P\left[\left\{ \sum_{j\in[M^{\beta}]}f^{+}\left(\tilde{G}(j)\right)\geq\E\left[\sum_{j\in[M^{\beta}]}f\left(\tilde{G}(j)\right)\right]+\eta\log M\right\} \cap{\cal F}^{c}\right]\\
 & \leq\P\left[\sum_{j\in[M^{\beta}]}f^{+}\left(\tilde{G}(j)\right)\geq\E\left[\sum_{j\in[M^{\beta}]}f\left(\tilde{G}(j)\right)\right]+\eta\log M\right].\label{eq: first upper bound on concentration of c}
\end{align}
The difference in expectation when switching from $f$ to $f^{+}$
is bounded, as in (\ref{eq: difference between expectations in f and f+-1}),
as 
\[
\sum_{j\in[M^{\beta}]}\E\left[f\left(\tilde{G}(j)\right)\right]-\E\left[f^{+}\left(\tilde{G}(j)\right)\right]\leq M^{\beta}e^{-M/2}\leq e^{-M/4}
\]
for all $M\geq M_{1}(P_{X},\beta)$. Hence, 
\begin{align}
 & \P\left[\left\{ \tilde{C}\geq\E[\tilde{C}]+\eta\log M\right\} \cap{\cal F}^{c}\right]\nonumber \\
 & \leq\P\left[\sum_{j\in[M^{\beta}]}f^{+}\left(\frac{\tilde{G}(j)}{M}\right)\geq\E\left[\sum_{j\in[M^{\beta}]}f^{+}\left(\frac{\tilde{G}(j)}{M}\right)\right]-e^{-M/4}+\eta\log M\right]\\
 & \leq\P\left[\sum_{j\in[M^{\beta}]}f^{+}\left(\frac{\tilde{G}(j)}{M}\right)\geq\E\left[\sum_{j\in[M^{\beta}]}f^{+}\left(\frac{\tilde{G}(j)}{M}\right)\right]+\frac{\eta}{2}\eta\log M\right]\label{eq: second upper bound on concentration of c}
\end{align}
for all $M\geq M_{2}(\eta)$. To further bound this probability, we
note that $\{f^{+}(\tilde{G}(j))\}_{j\in[M^{\beta}]}$ are IID RVs
which are bounded from above as
\[
f^{+}\left(\tilde{G}(j)\right)\leq\max_{t\geq0}(-t\log t)=\frac{1}{e}.
\]
We thus may use the regular Bernstein's inequality to bound the deviation
of their sum from its mean. To this end, we bound their second moment,
by noting that $\tilde{G}(j)\in\mathbb{N}_{+}$, and that for any
$g\in\mathbb{N}_{+}$ it holds that 
\[
0\leq f^{+}\left(g\right)\leq\frac{g}{M}\log M
\]
(as the intersection of the concave function $f(t)=-t\log t$ and
$t\log M$ occurs at $t=1/M$). Hence, 
\begin{align}
\E\left[\left(f^{+}\left(\tilde{G}(j)\right)\right)^{2}\right] & \leq\E\left[\left(\frac{\tilde{G}(j)}{M}\log M\right)^{2}\right]\\
 & =\frac{\log^{2}M}{M^{2}}\E\left[\tilde{G}^{2}(j)\right]\\
 & =\frac{\log^{2}M}{M}\P[X^{L}=a_{j}],
\end{align}
since $\tilde{G}(j)$ is Poisson with parameter $\E[\tilde{G}(j)]=M\cdot\P[X^{L}=a_{j}]$
for all $M\geq M_{0}$. So, Bernstein's inequality \cite[Corollary 2.11 and the discussion that follows it]{boucheron2013concentration}
implies that for any $r\geq0$
\begin{equation}
\P\left[\sum_{j\in[M^{\beta}]}f^{+}\left(\tilde{G}(j)\right)\geq\E\left[\sum_{j\in[M^{\beta}]}f^{+}\left(\tilde{G}(j)\right)\right]+r\right]\leq\exp\left[-\frac{r^{2}}{2\left(\frac{\log^{2}M}{M}+\frac{r}{3e}\right)}\right].\label{eq: Bernseteins bound on C concentration}
\end{equation}
Setting $r=\frac{\eta}{2}\log M$ in (\ref{eq: Bernseteins bound on C concentration})
we obtain 
\begin{align}
\frac{r^{2}}{2\left(\frac{\log^{2}M}{10M}+\frac{r}{3e}\right)} & \geq\frac{10Mr^{2}}{4\log^{2}M}\vee\frac{3e}{4}r\\
 & \geq\frac{1}{2}\cdot\left(\eta^{2}M\vee\eta\log M\right)\\
 & \geq\frac{1}{2}\cdot\eta\log M,
\end{align}
where the inequalities hold for all $M\geq M_{3}(\beta,\eta)$ large
enough. This, together with (\ref{eq: second upper bound on concentration of c})
and (\ref{eq: bound on C tilde concentration with F event}), implies
that 
\begin{align}
\P\left[\tilde{C}\geq\E[\tilde{C}]+\eta\log M\right] & \leq\exp\left[-\frac{1}{2}\eta\log M\right]+e^{-M\left(1-\frac{\beta\log M}{M}\right)}\\
 & \leq2\exp\left[-\frac{1}{2}\eta\log M\right]\\
 & =\frac{2}{M^{\eta/2}},
\end{align}
for all $M\geq M_{0}\vee M_{1}\vee M_{2}\vee M_{3}$. 
\end{IEEEproof}
We may now prove Theorem \ref{thm: Critical beta for repeat read regime}. 
\begin{IEEEproof}[Proof of Theorem \ref{thm: Critical beta for repeat read regime}]
If $H(P_{X})=0$ then trivially $\mathsf{FP}(\delta,\xi)=0$ for
any $\delta\geq0$ and $\xi\in[0,1]$. We thus assume henceforth that
$H(P_{X})>0$. In what follows, we will upper bound the pairwise error
probability between two sequences using the Bhattacharyya bound. To
this end, let $\tilde{x}^{N},\overline{x}^{N}\in{\cal X}^{N}$ be
a pair of sequences, where $\tilde{x}^{N}\neq\overline{x}^{N}$. Let
$p_{e}(\tilde{x}^{N}\to\overline{x}^{N})$ denote the error probability
of a pairwise test between $\tilde{x}^{N}$ and $\overline{x}^{N}$
from the observations $Y^{N}\sim P_{Y^{N}\mid X^{N}}(\cdot\mid\tilde{x}^{N})$.
Then, the Bhattacharyya bound on the probability of erroneously deciding
in favor of $\overline{x}^{N}$ in a pairwise test is given by (e.g.,
\cite[Sec. 2.3]{viterbi2009principles})
\begin{align}
p_{e}(\tilde{x}^{N}\to\overline{x}^{N}) & \leq\sum_{y^{N}\in{\cal Y}^{N}}\sqrt{P_{XY}^{\otimes N}[y^{N}\mid\tilde{x}^{N}]\cdot P_{XY}^{\otimes N}[y^{N}\mid\overline{x}^{N}]}\\
 & \trre[=,a]\sum_{y^{N}\in{\cal Y}^{N}}\prod_{i\in[M]}\sqrt{P_{XY}^{\otimes L}[\boldsymbol{y}(i)\mid\tilde{\boldsymbol{x}}(i)]\cdot P_{XY}^{\otimes L}[\boldsymbol{y}(i)\mid\overline{\boldsymbol{x}}(i)]}\\
 & =\prod_{i\in[M]}\sum_{\boldsymbol{y}\in{\cal Y}^{L}}\sqrt{P_{XY}^{\otimes L}[\boldsymbol{y}\mid\tilde{\boldsymbol{x}}(i)]\cdot P_{XY}^{\otimes L}[\boldsymbol{y}\mid\overline{\boldsymbol{x}}(i)]}\\
 & =e^{-\sum_{i\in[M]}d(\tilde{\boldsymbol{x}}(i),\overline{\boldsymbol{x}}(i))},\label{eq: Bhat basic pairwise error bound}
\end{align}
where $(a)$ holds since the fragments are independent. Assume that
$\delta>0$ and $\xi\in(0,1)$ are given, set $\eta\in(0,1)$, and
define the event 
\[
{\cal F}_{\eta}:=\left\{ x^{N}\in{\cal X}^{N}\colon\frac{1}{M}\log\left|{\cal A}_{L}(x^{N})\right|\leq L\cdot H(P_{X})+\eta\log M\right\} .
\]
Let us denote the failure error probability conditioned on $X^{N}=x^{N}$
by $\mathsf{FP}(\delta,\xi\mid x^{N})$. Then,
\begin{align}
\mathsf{FP}(\delta,\xi) & =\E\left[\mathsf{FP}(\delta,\xi\mid X^{N})\right]\\
 & \trre[\leq,a]\E\left[\mathsf{FP}(\delta,\xi\mid X^{N})\cdot\I\{X^{N}\in{\cal F}_{\eta}\}\right]+\P\left[X^{N}\not\in{\cal F}_{\eta}\right]\\
 & \trre[\leq,b]\E\left[\mathsf{FP}(\delta,\xi\mid X^{N})\cdot\I\{X^{N}\in{\cal F}_{\eta}\}\right]+o_{\eta}(1),\label{eq: failure prob upper bound  separation to bad and good events}
\end{align}
where $(a)$ follows from the union bound, and $(b)$ follows from
Prop. \ref{prop: number of possible reconstructions}. We next focus
on the first term. Let ${\cal S}_{L}(x^{N})\subset S_{M}$ be a set
of permutations that generates ${\cal A}_{L}(x^{N})$, that is, $|{\cal S}_{L}(x^{N})|=|{\cal A}_{L}(x^{N})|$
and for each $\tilde{x}^{N}\in{\cal A}_{L}(x^{N})$ there exists $\pi\in{\cal S}_{L}(x^{N})$
such that $\tilde{x}^{n}=\pi[x^{N}]$. Let $\mathsf{FP}(\delta,\xi,\pi[x^{N}]\mid X^{N}=x^{N})$
be the probability of the event in which the reconstruction failed
and the ML output was the erroneous $\pi[x^{N}]$. For any $x^{N}\in{\cal F}_{\eta}$
it then holds that 
\begin{align}
\mathsf{FP}(\delta,\xi\mid X^{N}=x^{N}) & \trre[\leq,a]\sum_{\pi\in{\cal S}_{L}(x^{N})}\mathsf{FP}(\delta,\xi,\pi[x^{N}]\mid X^{N}=x^{N})\\
 & \leq\left|{\cal S}_{L}(x^{N})\right|\cdot\max_{\pi\in{\cal S}_{L}(x^{N})}\mathsf{FP}(\delta,\xi,\pi[x^{N}]\mid X^{N}=x^{N})\\
 & \trre[\leq,b]e^{(\beta\cdot H(P_{X})+\eta)M\log M}\cdot\max_{\pi\in{\cal S}_{L}(x^{N})}\mathsf{FP}(\delta,\xi,\pi[x^{N}]\mid X^{N}=x^{N})\\
 & \trre[\leq,c]e^{(\beta\cdot H(P_{X})+\eta)M\log M}e^{-\xi M\cdot Ld^{*}(\delta)},\label{eq: failure prob for a given sequence and permutation}
\end{align}
where $(a)$ follows from the union bound, $(b)$ follows from Prop.
\ref{prop: number of possible reconstructions} and the assumption
that $x^{N}\in{\cal F}_{\eta}$, and $(c)$ follows from the following
consideration: Consider an arbitrary permutation $\pi\in{\cal S}_{L}(x^{N})$,
and denote $\tilde{x}^{n}=\pi[x^{N}]$. If 
\begin{equation}
\sum_{i\in[M]}\I\{\Delta(\boldsymbol{x}(i),\tilde{\boldsymbol{x}}(i))\geq\delta\}\geq\xi M\label{eq: the case in which the disortion is large for many reads}
\end{equation}
then the definition of $d^{*}(\delta)\equiv d_{P_{Y|X}}^{*}(\delta)$
in (\ref{eq: minimal Bhat for a given distortion}) implies that 
\[
\sum_{i\in[M]}d(\boldsymbol{x}(i),\tilde{\boldsymbol{x}}(i))\geq\xi ML\cdot d^{*}(\delta).
\]
In this case, (\ref{eq: Bhat basic pairwise error bound}) implies
that 
\[
\mathsf{FP}(\delta,\xi,\pi[x^{N}]\mid X^{N}=x^{N})\leq\exp\left[-\xi ML\cdot d^{*}(\delta)\right].
\]
Alternatively, if (\ref{eq: the case in which the disortion is large for many reads})
does not hold, we have that $\mathsf{FP}(\delta,\xi,\pi[x^{N}]\mid X^{N}=x^{N})=0$
(by the definition of reconstruction success at failure level $\xi$).
Inserting (\ref{eq: failure prob for a given sequence and permutation})
back to (\ref{eq: failure prob upper bound  separation to bad and good events}),
using $L=\beta\log M$, shows that if 
\[
\xi>\frac{H(P_{X})-\eta}{d^{*}(\delta)}
\]
then $\mathsf{FP}(\delta,\xi)=o_{\eta}(1)$ for all $M$ large enough.
The result then follows by taking  $\eta\downarrow0$.
\end{IEEEproof}
\bibliographystyle{plain}
\bibliography{DNA_ordering}

\end{document}